# The Effectiveness and Limits of Time-of-Use Pricing in Public EV Charging Networks

Mingzhi Xiao[1]*, Yuki Takayama[1]

**Abstract:** Time-of-use pricing is promoted to manage demand at public EV charging stations, yet its effectiveness depends on short run flexibility and local constraints. Using station by day by hour data from Shenzhen and Amsterdam, we estimate intraday price responsiveness on two margins, whether charging occurs in a station hour and, conditional on charging, delivered energy and occupancy time. High dimensional fixed effects absorb station by day demand shocks and hour of week patterns, so identification relies on within station, within day price variation under scheduled tariffs. Responses differ across cities. Shenzhen adjusts mainly through conditional intensity, whereas Amsterdam adjusts mainly through participation. Weather shifts responsiveness in opposite directions, with heat weakening responses in Shenzhen and rainfall strengthening participation responses in Amsterdam. Power upgrades typically outperform network densification except in transit-oriented areas.

## 1 Introduction

Transport electrification is moving fast in many cities. As the EV fleet grows, public charging becomes a daily service that can shape both grid operations and user experience. When and where drivers plug in affects peak load, local congestion at stations, and the need for expensive capacity upgrades. Because of this, time of use pricing is often promoted as a practical demand management tool. The basic idea is simple. Higher prices at stressed hours should push some charging to cheaper hours or nearby places, which can ease queues and reduce the need to build more hardware right away (Borlaug et al., 2023; He et al., 2022).

  This expectation relies on a strong behavioral assumption. It assumes that users have enough short run flexibility to change when they charge, where they charge, or how much they take. In practice, that flexibility is often limited. Home and workplace charging access, trip purpose, station availability, and local network layout can all constrain choices. When these constraints bind, prices may still matter, but the response may be small, or it may appear only in specific settings and hours. Reviews of charging behavior stress that price is only one part of a broader decision process that also depends on convenience, access, and situational limits (Potoglou et al., 2023; Shariatzadeh et al., 2025).

  Recent empirical work confirms that prices can move public charging outcomes, including energy delivered and station choice. At the same time, the evidence also shows that responses are uneven across places and conditions. Non price factors such as availability, waiting, and network structure can weaken or reshape observed price responsiveness even under well designed schedules (Kuang et al., 2024; Yang et al., 2024). Weather and operating conditions add another layer. Cold or adverse conditions can reduce flexibility and change how users value time and comfort, which can change how strongly charging shifts across hours (Bayram & Galloway, 2021). These findings suggest that average elasticities can hide important heterogeneity across time, space, and states.

---

[1] Institute of Science Tokyo, 2-12-1 W6-9, Ookayama, Meguro-ku, Tokyo 152-8552, Japan
\* Corresponding author: xiao.m.1475@m.isct.ac.jp

This paper focuses on a basic but practical question. Under real time of use regimes in public networks, when does price have leverage, and through what short run channel does it operate. We emphasize two margins that can look similar in aggregate but imply very different operational levers. The first is the participation margin, which asks whether charging occurs in a station hour. The second is the conditional intensity margin, which asks how much energy is delivered and how long chargers are occupied once charging occurs. This distinction matters because pricing can reduce the number of sessions in some settings, while in other settings it mainly changes within session behavior, such as how long a car stays plugged in or how much energy is taken (Potoglou et al., 2023; Shariatzadeh et al., 2025).

We study these issues using high frequency station by day by hour data from two contrasting urban systems, Shenzhen and Amsterdam. We combine charging outcomes with posted total prices that include electricity tariffs and service fees. The identification strategy uses rich fixed effects and exploits within station and day price variation induced by administratively scheduled time of use regimes. Station by day fixed effects absorb local daily shocks such as outages or localized demand surges, while hour of week fixed effects absorb systematic weekly patterns. This design targets short run intraday responses within the pricing schedules that users actually face.

The cross city setting is not only a comparison of levels. It is a way to learn how institutions, network form, and user constraints change the binding margin of adjustment. In some contexts, public charging may be closer to a necessary activity with limited timing options, so price may mainly change what happens after a session starts. In other contexts, charging may be more discretionary, so price may mainly change whether a session starts at all. Weather and land use context can shift these margins by changing flexibility, travel patterns, and the cost of search and waiting (Bayram & Galloway, 2021; Shariatzadeh et al., 2025).

Beyond the main panel analysis, we also connect short run estimates to practical planning questions. Using cross sectional station level patterns, we study how effective capacity and very local station density relate to per station throughput. The goal is not to claim a welfare optimum. Instead, we use these relationships to rank operational levers and to diagnose which constraint is more likely to bind in different micro environments. Related work highlights that congestion, waiting, and network feedback can matter for observed demand responses, and that equity and access constraints shape who relies on public networks (Huang & Kockelman, 2020; Yu et al., 2025). These issues motivate careful interpretation and robustness checks, and they also motivate the counterfactual scenarios that compare upgrades with ultra local densification.

This paper contributes in three ways. First, it provides high resolution evidence on intraday price responsiveness in public charging using two cities and a unified data structure. Second, it separates participation from conditional intensity, which clarifies why similar average elasticities can reflect different adjustment channels. Third, it shows how operating states and spatial context shape where pricing works best and where non price constraints likely dominate. These results translate into simple operational guidance on when to prioritize pricing, when to prioritize upgrades, and when to be cautious about adding stations very close to each other.

The rest of the paper is organized as follows. Section 2 reviews related literature. Section 3 describes the data and variable construction. Section 4 presents the empirical strategy, main results, robustness checks, and policy counterfactuals. Section 5 discusses interpretation and practical implications. Section 6 concludes.

## 2 Literature Review

Public EV charging is studied from two closely related angles. One angle treats it as an infrastructure and operations problem. The other treats it as a user choice problem. On the infrastructure side, many studies show that the need for public charging and its performance depend on context. In dense areas, public charging can substitute for limited home charging, while in other places it plays more of a backup role. This means that simple coverage targets can be misleading, and planning decisions should consider how charging demand is distributed across time and space (Funke & Sprei, 2019). Related work also highlights that policy choices often involve a tradeoff between investing in more public fast-charging and investing in other forms of flexibility, such as longer ranges or different charging access, and that the "best" option depends on behavior and travel patterns (Funke et al., 2019). At the same time, station-level utilization is far from uniform. Even within the same city, some stations operate close to capacity while others remain underused, which makes network management and pricing relevant to daily operations rather than only to long-run expansion decisions (Wolbertus & Gerzon, 2018). These facts motivate demand-side tools that can reallocate charging across hours or across nearby stations, instead of relying only on adding hardware.

On the behavior side, the core idea is that charging decisions are constrained choices. Users face not only money costs but also time costs, search costs, and schedule constraints. Standard scheduling and congestion ideas imply that people may shift activities across time when prices change, but only if they can absorb schedule delay or change their routine (Small, 1982; Vickrey, 1969). In public charging, this suggests that short-run responses to price should mainly show up as intraday adjustments, such as moving charging to cheaper hours or choosing a different station nearby, while larger adjustments, such as changing where to live or relying more on home charging, are longer-run decisions. Empirical and review studies support this view. A recent review on public charging choices stresses that observed charging behavior is shaped by a mix of price, convenience, access, and situational constraints, and that user heterogeneity makes average effects hard to interpret without looking at context (Potoglou et al., 2023). A related literature on "smart charging" also emphasizes that pricing interacts with activity schedules and constraints, so a realistic response model needs to consider the joint nature of travel, parking, and charging (Daina et al., 2017). More broadly, evidence from dynamic electricity pricing shows that consumers do respond to time-varying prices, but response sizes depend strongly on program design, information, and the constraints users face (Faruqui & Sergici, 2010). This body of work implies that TOU pricing can work in public charging, but the response is unlikely to be stable across settings.

Recent charging-specific studies provide direct evidence that posted prices matter, while also highlighting limits. For example, a study using public charging data from Shenzhen estimates that demand is price inelastic on average, and it documents meaningful spatial and temporal variation in responsiveness, which suggests that prices have leverage in some situations but less so in others (Kuang et al., 2024). At the same time, work that focuses on station efficiency shows that non-price frictions such as local parking pressure and user routines can shape how effective a pricing fee is in improving turnover, and the response can differ across drivers and neighborhoods (Wolbertus & Gerzon, 2018).

Network interactions further complicate interpretation. When multiple stations are close substitutes, a price change at one location can shift demand within a local market, and congestion and waiting can feed back into both route choice and station choice. A network equilibrium model that explicitly includes station congestion and elastic charging demand shows that these feedbacks can be important and can change the observed price response (Huang & Kockelman, 2020). These insights mean that reduced-form estimates of price responsiveness are still valuable for operations, but they should be read as outcomes of both preference and constraints, and not as a complete explanation of the behavioral mechanisms.

This paper builds on the literature above but targets a specific operational object that is still not well documented. We focus on short-run intraday responses under real posted TOU regimes in public networks. We use high-frequency station-by-day-by-hour data and distinguish two margins through which price can operate. The first is the participation margin, which asks whether any charging occurs in a station-hour. The second is the conditional intensity margin, which asks how much energy is delivered and how long chargers are occupied once charging occurs. This distinction is important because similar average elasticities can reflect different adjustment channels, and these channels imply different operational responses. We then examine how these price responses vary with weather and place context. The motivation follows directly from prior work. If flexibility and constraints vary across time, weather, and local environments, then the effectiveness of pricing should vary as well (Potoglou et al., 2023; Daina et al., 2017). Finally, by comparing Shenzhen and Amsterdam, we aim to show how similar policy tools can operate differently across systems, and we link these findings to practical planning levers such as upgrades versus very local densification, while acknowledging that reduced-form responses cannot fully identify the exact mechanism behind each change without richer user-level or queueing data.

## 3 Data and Methods

### 3.1 Data

We assemble a harmonized station by day by hour panel for Shenzhen (SZ) and Amsterdam (AMS) by linking hourly operational records from public charging networks to posted tariffs and local weather. The underlying data come from a standardized global EV charging dataset (Guo et al., 2025). This dataset provides city scale and harmonized station level observations that allow for consistent cross city analysis.

For each station $i$, day $d$, and hour $h$, the data include energy delivered $q_{idh}$ in kWh, occupancy duration $dur_{idh}$ in hours, and the total retail price per kWh defined as energy tariff plus service fee. These variables are matched to hourly temperature $T_{idh}$ in degrees Celsius and precipitation $R_{idh}$. Timestamps are converted to local civil time and mapped to an hour of week index from 0 to 167. Daylight saving time transitions are handled explicitly for Amsterdam. The two cities provide complementary operating environments.

In Shenzhen, station hour activity is frequent and right skewed. A small set of hours accounts for a disproportionate share of energy and time. Tariffs exhibit coarse steps and a visible mass near zero that reflects fee waivers or strongly discounted off peak windows, with wide intraday spreads under time of use schedules. The climate is warm and humid with recurrent hot hours and episodic heavy rainfall. These features imply many hours with strong utilization and substantial within day variation in prices. In Amsterdam, usage is sparser but more regular. A sizable fraction of station hours record no activity and positive hours cluster around moderate intensities. Prices are dominated by the energy component with limited-service fee dispersion and narrower intraday spreads. The temperate and rain prone climate yields lower average temperatures. These conditions generate more situations where the key decision is whether to initiate a session rather than how much to charge once started.

For the hourly analyses in RQ1 and RQ2, we report city specific descriptive statistics for Energy, Duration, Energy price, Service fee, Total price, Temperature, and Precipitation in Table 1 for Shenzhen and Table 2 for Amsterdam. These tables document distributional shape such as masses at zero and the scale of price and weather variation prior to estimation. For the station level analyses in RQ3 and RQ4, we aggregate over the observation window to form a cross section by site and report Total energy, Total occupancy time, Energy per charger, Time per charger, Number of chargers, and Average rated power. These outcomes cover both levels and per charger intensities and help separate capacity from utilization. They appear in Table 3 for Shenzhen and Table 4 for Amsterdam. In keeping with the descriptive scope of this section and current data availability, distance-based network measures are not included in these tables. Amsterdam lacks validated between station distances and Shenzhen distance metrics are omitted here by design.

To ensure cross city comparability, variable names, units, and construction are harmonized. Prices that enter logs are required to be strictly positive and a small constant is added where log transforms are applied to outcomes with zeros. Continuous covariates are winsorized at conventional tails. These features create semi continuous outcomes with mass at zero and substantial heterogeneity in weather and tariffs. They motivate the two-part specification for RQ1 and the price by weather interactions for RQ2 that are detailed in the empirical strategy. Beyond what is tabulated here, the dataset also supports the geospatial descriptors introduced in Figure 1 such as land use and points of interest context, which we leverage later when modeling station level heterogeneity in RQ3 and when scaling RQ4 counterfactuals. These additions do not alter the descriptive tables in this subsection.

**Table 1. Descriptive statistics (Shenzhen, RQ1-RQ2)**

| Variable (units) | N | Mean | SD | P10 | P25 | Median | P75 | P90 | Min | Max |
|---|---|---|---|---|---|---|---|---|---|---|
| Energy (kWh) | 6,346,440 | 93.80 | 502.22 | 0.00 | 4.38 | 12.83 | 43.24 | 134.82 | 0.000 | 16,006.67 |
| Duration (hours) | 6,346,440 | 5.37 | 13.11 | 0.32 | 0.67 | 1.75 | 4.75 | 13.33 | 0.000 | 439.65 |
| Energy price (CNY/kWh) | 6,346,440 | 0.188 | 0.356 | 0.000 | 0.000 | 0.000 | 0.290 | 0.876 | 0.000 | 1.500 |
| Service fee (CNY/kWh) | 6,346,440 | 0.100 | 0.236 | 0.000 | 0.000 | 0.000 | 0.000 | 0.700 | 0.000 | 0.800 |
| Total tariff (CNY/kWh) | 6,346,440 | 0.288 | 0.520 | 0.000 | 0.000 | 0.000 | 0.290 | 1.100 | 0.000 | 2.060 |
| Temperature (°C) | 6,346,440 | 28.0 | 3.2 | 23.5 | 26.1 | 28.5 | 30.0 | 31.8 | 17.6 | 37.0 |
| Precipitation (mm/h) | 6,346,440 | 0.26 | 3.10 | 0.00 | 0.00 | 0.00 | 0.00 | 0.00 | 0.00 | 107.51 |

**Table 2. descriptive statistics (Amsterdam, RQ1-RQ2)**

| Variable | N | Mean | SD | P10 | P25 | Median | P75 | P90 | Min | Max |
|---|---|---|---|---|---|---|---|---|---|---|
| Energy (kWh) | 10756008 | 13.92 | 63.79 | 0.00 | 0.00 | 0.00 | 6.93 | 19.90 | 0.00 | 3789.51 |
| Duration (hours) | 10756008 | 1.71 | 5.38 | 0.00 | 0.00 | 0.00 | 1.08 | 3.00 | 0.00 | 153.42 |
| Energy price (€/kWh) | 10756008 | 0.26 | 0.20 | 0.00 | 0.00 | 0.31 | 0.39 | 0.42 | 0.00 | 0.91 |
| Service fee (€/kWh) | 10756008 | 0.00 | 0.04 | 0.00 | 0.00 | 0.00 | 0.00 | 0.00 | 0.00 | 1.00 |
| Total price (€/kWh) | 10756008 | 0.26 | 0.21 | 0.00 | 0.00 | 0.31 | 0.39 | 0.42 | 0.00 | 1.76 |
| Temperature (°C) | 10756008 | 15.83 | 5.12 | 8.80 | 12.30 | 16.20 | 19.10 | 21.90 | −0.10 | 31.20 |
| Precipitation (mm/h) | 10756008 | 0.09 | 0.65 | 0.00 | 0.00 | 0.00 | 0.00 | 0.00 | 0.00 | 19.44 |

**Table 3. Descriptive statistics (Shenzhen, RQ3-RQ4)**

| Variable (units) | N | Mean | SD | P25 | Median | P75 |
|---|---|---|---|---|---|---|
| Total energy delivered (kWh) | 1,445 | 1,335,266 | 17,988,513 | 41,514 | 108,289 | 313,362 |
| Total occupancy time (hours) | 1,445 | 45,077 | 104,376 | 6,966 | 15,454 | 42,914 |
| Energy per charger (kWh/charger) | 1,445 | 819,233 | 6,729,507 | 35,770 | 75,282 | 214,620 |
| Time per charger (h/charger) | 1,445 | 29,479 | 55,113 | 5,192 | 10,993 | 27,891 |
| Number of chargers (count) | 1,445 | 1.519 | 0.614 | 1 | 1 | 2 |
| Average rated power (kW) | 1,445 | 13.078 | 31.649 | 7 | 7 | 7 |

**Table 4. descriptive statistics (Amsterdam, RQ3-RQ4)**

| Variable (units) | N | Mean | SD | P25 | Median | P75 |
|---|---|---|---|---|---|---|
| Total energy delivered (kWh) | 2,449 | 97,662.471 | 404,068.179 | 0.000 | 30,731.200 | 62,790.763 |
| Total occupancy time (hours) | 2,449 | 11,763.338 | 31,681.543 | 0.000 | 4,761.292 | 9,811.057 |
| Energy per charger (kWh/charger) | 2,449 | 58,742.417 | 153,103.123 | 0.000 | 17,071.733 | 53,829.591 |
| Time per charger (h/charger) | 2,449 | 7,423.431 | 15,926.319 | 0.000 | 2,638.708 | 8,389.113 |
| Number of chargers (count) | 2,449 | 1.440 | 0.982 | 1.000 | 1.000 | 2.000 |
| Average rated power (kW) | 2,449 | 3.470 | 3.597 | 0.000 | 3.200 | 6.400 |

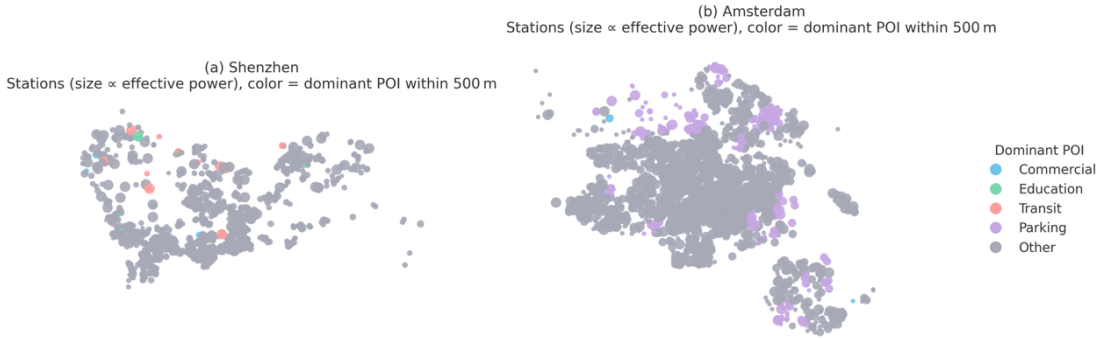

**Figure 1. Spatial context of public charging stations.**

## 3.2 Empirical framework and notation

Let $c \in \{SZ, AMS\}$ index city, $i$ station, $d$ day, and $h$ hour. We study two behavioral margins:
- the extensive margin, whether any charging occurs in a station–hour, proxied by the binary

indicators

$$ExtDur_{cidh} \equiv 1\{dur_{cidh} > 0\}, \qquad ExtEn_{cidh} \equiv 1\{q_{cidh} > 0\};$$

These two indicators capture participation under two measurement channels, occupancy time and delivered energy, and provide a consistency check when one metric is missing or noisier.

- the intensive margin, the conditional magnitude of charging once it occurs, captured by

$$ln\,(dur_{cidh} + \epsilon) \text{ and } ln\,(q_{cidh} + \epsilon).$$

where $\epsilon$ is a small constant to accommodate zero observations.

We exploit within-station, within-day variation across hours in posted total tariffs $p_{cidh}$. Hourly tariffs in both cities follow administratively scheduled time-of-use regimes and are predetermined at daily or longer horizons. Prices are not adjusted in response to contemporaneous station congestion or realized hourly demand. As a result, identifying variation arises from within-day tariff schedules rather than from real-time pricing responses, supporting an interpretation in terms of short-run intraday charging decisions.

Figure 2 plots average posted total prices (energy tariff plus service fee) by hour-of-week in each city (CNY/kWh for Shenzhen and EUR/kWh for Amsterdam). It shows wider intraday dispersion in Shenzhen and a flatter profile in Amsterdam, which motivates our focus on hourly intraday price responses in RQ1.

Identification in our main specifications combines the exogeneity of scheduled tariffs with high-dimensional fixed effects. All models include station×day fixed effects ($\alpha_{id}$) to absorb station-specific demand shifters that are constant within a day, such as local events, temporary access constraints, or unobserved day-level surges in demand. To purge common weekly periodicity in charging activity and pricing schedules, we also include hour-of-week fixed effects ($\tau_{h(w)}$). Under this structure, identification relies on the remaining within-station, within-day variation across hours that is not explained by station-by-day demand conditions or by common hour-of-week patterns. In our data, posted total prices combine an energy tariff and a service fee, and they may differ across operators and stations, with occasional within–hour-of-week deviations due to tariff components and implementation rules. We therefore interpret the estimated price response as being identified from intraday price differences that are plausibly orthogonal to short-run station-hour demand shocks, conditional on ($\alpha_{id}$, $\tau_{h(w)}$).

Standard errors are clustered at the station (pile) level. For interpretability across cities, temperature is mean centered within city so that $T_{cidh} = 0$ represents typical conditions. Rainfall is coded as an indicator in the baseline and as intensity in robustness checks.

This framework allows us to quantify intraday price responsiveness and then examine how it varies with weather and spatial context. We pursue four research questions:

- RQ1: short-run price elasticity on extensive and intensive margins.
- RQ2: weather-dependent moderation of the price response (temperature, rainfall).
- RQ3: station-level spatial heterogeneity in utilization.
- RQ4: counterfactual simulations for tariff and climate scenarios.

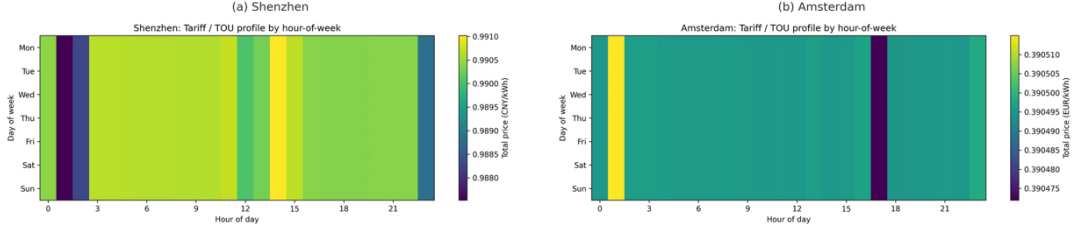

**Figure 2. Hour-of-week tariff profiles**

### 3.3 Intraday price elasticity (two-part design) (RQ1)

Because station-hour outcomes are zero-inflated, we estimate a two-part model that separates participation from conditional intensity. We implement the design with high-dimensional fixed effects, so identification comes from within-station, within-day comparisons across hours after absorbing station-day demand shocks and common hour-of-week patterns.

Extensive margin (linear probability model). For $Y \in \{ExtDur, ExtEn\}$,

$$Y_{cidh} = \beta_0 + \beta_1 \ln(p_{cidh}) + \alpha_{id} + \tau_{h(w)} + \varepsilon_{cidh}$$

(3.3.1)

In (3.3.1), β1\beta_1β1 is a semi-elasticity of participation with respect to price. A 10% increase in price implies an approximate change in the charging probability of

$$\Delta Y \approx \beta_1 \cdot \ln(1.10),$$

(3.3.1a)

reported in percentage points for interpretability.

Intensive margin (log–log conditional model). For $Z \in \{dur, q\}$,

$$\ln(Z_{cidh} + \epsilon) = \gamma_0 + \gamma_1 \ln(p_{cidh}) + \alpha_{id} + \tau_{h(w)} + u_{cidh}, \quad on\ Z_{cidh} > 0$$

(3.3.2)

In (3.3.2), $\gamma_1$ is the semi-elasticity of charging intensity with respect to price: a 10% increase in price maps to

$$\Delta \ln Z \approx \gamma_1 \cdot \ln(1.10)$$

(3.3.3)

so $\gamma_1 < 0$ indicates that higher prices reduce conditional minutes or delivered energy, holding fixed $(\alpha_{id},\ \tau_{h(w)})$.

Alongside the log specifications for relative effects, we also estimate OLS in levels for absolute policy translation (e.g., minutes or kWh per +0.10 currency/kWh). These level models combine extensive and intensive changes, and we do not treat them as headline elasticities.

All RQ1 specifications are estimated separately by city. Pooled variants include interactions between $\ln p$ and a city indicator to test for heterogeneous elasticities across Shenzhen and Amsterdam. Standard errors are clustered at the station (pile) level.

### 3.4 Weather-state dependence of price responsiveness (RQ2)

To allow the price response to vary with ambient conditions, we interact $\ln p$ with temperature and rainfall. The extensive-margin specification augments (3.3.1) as

$$Y_{cidh} = \beta_0 + \beta_1 \ln p_{cidh} + \beta_2 (\ln p_{cidh} \times T_{cidh}) + \beta_3 (\ln p_{cidh} \times R_{cidh}) + \alpha_{id} + \tau_{h(w)} + \varepsilon_{cidh}$$

(3.4.1)

and the intensive-margin specification augments (3.3.2) as

$$\ln(Z_{cidh} + \epsilon) = \gamma_0 + \gamma_1 \ln p_{cidh} + \gamma_2 (\ln p_{cidh} \times T_{cidh}) + \gamma_3 (\ln p_{cidh} \times R_{cidh}) + \alpha_{id} + \tau_{h(w)} + \varepsilon_{cidh}$$

(3.4.2)

Under either margin, the effective price slope at hour $h$ is the combination of main and interaction terms,

$$\left.\frac{\partial \ln Z}{\partial \ln p}\right|_{cidh} = \gamma_1 + \gamma_2 T_{cidh} + \gamma_3 R_{cidh}$$

$$(or\ \beta_1 + \beta_2 T_{cidh} + \beta_3 R_{cidh}\ for\ the\ LPM)$$

(3.4.3)

so that a 10% increase in price implies

$$\Delta \ln Z \approx \gamma_1 + \gamma_2 T_{cidh} + \gamma_3 R_{cidh}$$

(3.4.4)

evaluated at the weather state $(T, R)$ of interest. In the results (Section 4.2), we summarize these as average marginal effects (AMEs) at representative conditions, the city-specific median temperature and rainy vs. non-rainy hours, so readers can interpret weather-conditioned elasticities without revisiting the derivation.

Two remarks guide interpretation. First, because $\alpha_{id}$ and $\tau_{h(w)}$ absorb station–day shocks and weekly periodicity, the main price coefficient alone need not align with casual intuition; the economically relevant object is the effective slope (3.4.3). Second, centering temperature implies that $\gamma_1$ (or $\beta_1$) is the price slope at typical conditions, while $\gamma_2$, $\gamma_3$ (or $\beta_2$, $\beta_3$) describe how heat and rain tilt that slope.

## 3.5 Data aggregation, outcomes, and covariates (harmonized across cities)

We study public EV charging at two complementary resolutions aligned with the research questions (RQs). RQ1–RQ2 analyze hourly station–time panels to identify temporal responses (e.g., TOU/price, peak vs. off-peak, weather). RQ3–RQ4 analyze the cross-section fixed over the observation window to explain spatial heterogeneity and to project policy counterfactuals.

For the cross-section (RQ3–RQ4), the unit is the charger. The primary dependent variable is $log\ y_{ic}$, the log of total energy (kWh) per charger $i$ at station sss over the sample window. Robustness outcomes include occupied time per charger and $log\ (y_{ic} + 1)$ to accommodate zeros. For the panel (RQ1–RQ2), we use station–hour energy (and sessions), retaining zeros to avoid selection.

We harmonize the following across cities:
(i) Rated power $P_{ic}$ (kW), entered as $log\ P_{ic}$;
(ii) Station size $S_s$ (number of connectors on the site), $log\ S_s$;
(iii) Nearest-neighbor spacing between stations (km), $log\ NND_s$, computed from coordinates;
(iv) Very-near competition within 0.5 km, $log(1 + c_{0.5km,s})$, where $c_{0.5km,s}$ counts other public stations within 500 m. These separate capacity (power, size) from ultra-local crowding (spacing, very-near competitors).

Each station is assigned a dominant POI background within 500 m using OSM-type POIs aggregated to: commercial (food/retail), education, health, transit, parking, leisure, civic, and other. To maintain power, we merge sparse categories when needed (e.g., Amsterdam reduces to parking vs other in estimation). Background enters as fixed effects and, where relevant, in interactions with competition.

Variable construction follows a single protocol in both cities. Amsterdam does not include a separate "distance dataset" in descriptive tables; density/competition measures are derived from coordinates. Shenzhen has additional distance covariates that we keep for robustness only, to preserve comparability in the main specifications. Sample definitions, trimming rules, and winsorization thresholds are shared and reported alongside city Ns in §4.3–§4.4.

## 3.6 RQ3: Cross-sectional specification and estimation

We explain cross-sectional differences in realized throughput per charger using equipment, network structure, and local background.

Baseline model

$$log\ y_{ic} = \alpha + \beta_1 log\ P_{ic} + \beta_2 log\ S_s + \beta_3 log\ NND_s + \beta_4 log(1 + c_{0.5km,s}) + \gamma_{g(s)} + \varepsilon_{ic}$$
(3.6.1)

where $g(s)$ is the station's dominant 500 m background (baseline "other"). We estimate by OLS with standard errors clustered at the station level. Coefficients $\beta_1$ and $\beta_4$ are interpreted as conditional power and very-near competition elasticities.

To allow crowding effects to vary by context, we interact competition with background:

$$log\ y_{ic} = \alpha + \beta_1 log\ P_{ic} + \beta_2 log\ S_s + \beta_3 log\ NND_s + \beta_4 log(1 + c_{0.5km,s}) + \sum_{g \in G\setminus\{other\}} \delta_g D_{sg}$$
$$+ \sum_{g \in G\setminus\{other\}} \theta_g [D_{sg} \times log(1 + c_{0.5km,s})] + \varepsilon_{ic}$$
(3.6.2)

This delivers a competition elasticity for "other" ($\beta_4$) and background-specific elasticities $\beta_4 + \theta_g$.

We trim impossible values, inspect leverage (Cook's D), and perform leave-one-station / leave-one-operator jackknife. Given spatial correlation, we supplement clustered SEs with Conley / neighborhood-cluster SEs. We also vary competition definitions (0.3/0.5/1.0 km; distance-weighted sums; kNN with 5/10 nearest) and background coding (dominance thresholds 40–50%; merging small classes). Heavy-tail robustness uses winsorization and median-based outcomes. Main results are presented by city in §4.3.

## 3.7 RQ4: Counterfactual mapping and scenario design

RQ4 converts RQ3 elasticities into associational policy counterfactuals over the observed cross-section.

For a power shock $\Delta P$: $y_i^{cf} = y_i \times (1 + \hat{\varepsilon}_P \Delta P)$.

For a +1 competitor within 0.5 km: $y_i^{cf} = y_i \times exp\ (\hat{\varepsilon}_{comp}^g \cdot \Delta log(1 + c))$, with $\Delta log\ (1 + c) = log\frac{c+2}{c+1}$ and $\hat{\varepsilon}_{comp}^g$ taken from the background-specific RQ3 estimates. We aggregate counterfactual $y_i^{cf}$ to city totals and background subtotals, verifying that background deltas sum to the city delta.

Scenarios (harmonized across cities; targets city-specific):

S1: +10% power for all chargers.

S2: +50% power for low-power chargers (bottom quartile). In Amsterdam we define the bottom quartile by rank among positive-power chargers (to avoid 6.4 kW mass points).

S3/S4: +1 very-near competitor targeted to a background (Shenzhen: transit / education; Amsterdam: parking / other).

S5: S2 then S3, multiplicatively (upgrade first, then targeted densification).

Scenario outcomes for each city are reported in §4.4 (totals and reallocations).

## 3.8 Inference, diagnostics, and robustness (RQ1–RQ4)

All baseline specifications report heteroskedasticity robust standard errors clustered at the station level. Where feasible, we also provide alternative inference using wild cluster procedures or date-based clustering as a reference. These alternatives are consistent with the station clustered results and do not change interpretation. We treat station clustered inference as primary.

RQ1–RQ2 (hourly panel). We estimate price/TOU effects with high-dimensional fixed effects (station×hour-of-day; station×day-of-week; date or station×date), weather controls (levels and peak/off-peak interactions), and holiday dummies. Inference uses station-clustered SEs; key coefficients receive wild-cluster/bootstrap checks. Functional-form robustness compares OLS with PPML (handles zeros/skew) and a two-part model (incidence × intensity). Placebo tests include lead terms of tariffs/prices (expected null). Sub-samples (weekday/weekend; peak/off-peak; rain/clear; power bands) document heterogeneity consistent with §4.1–§4.2 findings.

RQ3 (cross-section). Main inference uses station-clustered SEs; robustness adds Conley / neighborhood-cluster SEs to address spatial correlation. We vary competition metrics (radius, distance weights, kNN), background coding, and outcome definitions (winsorized / median-based). Jackknife (by station/operator) and permutation tests (randomized neighborhood counts within districts) guard against leverage and spurious spatial correlation. These checks support the patterns summarized in §4.3.1 (Shenzhen) and §4.3.2 (Amsterdam): steep power elasticities; context-specific competition effects.

RQ4 (counterfactuals). Because RQ4 applies point estimates from RQ3, we propagate uncertainty by parametric simulation of the RQ3 parameter vector (≥1,000 draws from the estimated multivariate normal), recomputing S1–S5 each time. We report 95% confidence intervals for citywide deltas in the main text and for background deltas in the appendix. We also perform definition sensitivity (S2 threshold at 20/25/30%; competition radius 0.3/0.5/1.0 km; ±1 competitor to contrast densification vs. de-densification) and accounting checks (background deltas sum to city delta). For policy readability, we present a cost-effectiveness frontier (ΔkWh per unit CAPEX) ranking upgrades vs. densification, and a transportability check that projects each city using the other city's elasticities to highlight corridor-vs-convenience regimes (§4.4.3).

Interpretation. Throughout, coefficients and counterfactuals are interpreted as conditional associations under observed structure. RQ1–RQ2 document temporal responses that inform the demand environment; RQ3 explains cross-sectional utilization given equipment and local crowding; RQ4 ranks policy levers under those elasticities. Together, the evidence underpins the expansion logic used in §4.4: upgrade-first, with context-gated densification (corridor-like hubs in Shenzhen; minimum separations and share caps in Amsterdam's parking-dominant micro-areas).

# 4 Analysis Results

The results are organized around the four research questions introduced above. Unless otherwise noted, we treat the two-part specification with station×day fixed effects and hour of week fixed effects as the primary specification. This design absorbs station specific day level demand shocks and common weekly timing patterns, so the estimated price responses are identified from within station within day variation across hours that remains after these controls. Specifications with weaker or alternative fixed effects are reported as sensitivity checks and help gauge how conclusions change when the control structure is relaxed. All elasticities reported below capture short run intraday adjustments in charging behavior and should be interpreted as operational responses to posted prices under observed time of use regimes, rather than as long run demand elasticities.

### 4.1 RQ1: Price responsiveness of public EV charging

This section documents how the posted total retail price of public charging (energy tariff plus service fee) is related to charging activity at two margins: participation (the extensive margin) and conditional magnitude (the intensive margin). Because station–hour outcomes are zero-inflated, we use a two-part design that separates whether any charging occurs from how much charging occurs conditional on activity. We report three complementary families of models: (i) OLS in levels and log–log form, (ii) high-dimensional fixed effects (HDFE) with station×day and 168 hour-of-week dummies, and (iii) a two-part design that explicitly separates extensive from intensive responses. For readability across audiences, log–log coefficients are translated into percent changes for a 10% price increase, while levels coefficients are translated into absolute changes for a +0.10 local-currency/kWh step. Standard errors are HC1 for OLS and clustered at the station (pile) level for all fixed-effects regressions. Full estimates appear in Table 5 (Shenzhen) and Table 6 (Amsterdam). Figures 3–6 visualize hour-of-week patterns of responsiveness, and Figure 7 decomposes the response to a +10% price change into extensive (participation) and intensive (conditional kWh) components.

Throughout, we interpret the estimates as conditional associations and short-run intraday responsiveness under observed time-of-use regimes, rather than as definitive structural causal effects. The richer fixed-effects specifications sharpen a within-station, within-day interpretation by absorbing station-day shocks and common hour-of-week timing patterns, but unobserved implementation details or operator policies may still co-move with posted prices.

We use two operational definitions of the extensive margin: Extensive-duration $\equiv \mathbb{1}\{\text{duration} > 0\}$ and Extensive-energy $\equiv \mathbb{1}\{\text{kWh} > 0\}$. In practice they coincide in many hours; any differences arise from occasional measurement noise or rounding at very low values. Unless otherwise noted, extensive-margin effects are expressed relative to each indicator's mean participation rate (see Tables 1–2), which makes participation responses comparable across indicators and cities.

### 4.1.1 Shenzhen

In Shenzhen, the two-part specification without fixed effects shows economically modest but precisely estimated negative price responsiveness at both margins (Table 5, Panel C: Two-part (No-FE)). A 10% price increase is associated with a small reduction in the probability of positive activity, and with modest reductions in conditional duration and conditional delivered energy among active hours.

When station×day and hour-of-week fixed effects are included in the two-part design (Table 5, Panel C: Two-part (FE)), extensive-margin coefficients move close to zero and become statistically indistinguishable from zero. The conditional duration coefficient remains negative and statistically significant, while the conditional kWh coefficient is imprecisely estimated. This pattern is consistent with the idea that, under a within-station, within-day interpretation, Shenzhen's intraday price responsiveness is more visible in conditional time spent charging than in participation, but the magnitude can be sensitive when identification relies on the residual price variation left after strong fixed effects.

OLS in levels and log–log form provides an additional descriptive summary, but it mechanically mixes the extensive and intensive margins and can reflect compositional differences, so we treat it as complementary context rather than a headline measure of responsiveness.

**Table 5. Price–Charging Response (Shenzhen, RQ1)**

| Panel | Outcome | Individual FE | Time FE | Coef (SE) | p | N |
|---|---|---|---|---|---|---|
| **A. No-FE (OLS, HC1)** | Duration (hours) | | | -37.343 (0.358) | 0.000*** | 204,960 |
| | Energy (kWh) | | | -291.592 (1.875) | 0.000*** | 11,712 |
| | ln (Duration+ε) | | | -2.705 (0.022) | 0.000*** | 204,960 |
| | ln (kWh+ε) | | | -1.443 (0.006) | 0.000*** | 11,712 |
| **B. FE (station×day + 168h)** | Duration (hours) | YES | YES | 699.749 (191.519) | 0.000*** | 204,960 |
| | Energy (kWh) | YES | YES | -84.813(7158.505) | 0.991 | 11,712 |
| | ln (Duration+ε) | YES | YES | -30.160 (15.571) | 0.053* | 204,960 |
| | ln (kWh+ε) | YES | YES | -25.066 (5.954) | 0.000*** | 11,712 |
| **C. Two-part (No-FE)** | 1{Duration>0} | | | -0.024 (0.000) | 0.000*** | 1,845,189 |
| | 1{kWh>0} | | | -0.014 (0.000) | 0.000*** | 1,845,189 |
| | ln (Duration) \| Duration>0 | | | -0.091 (0.002) | 0.000*** | 1,646,771 |
| | ln (kWh) \| kWh>0 | | | -0.274 (0.004) | 0.000*** | 106,995 |
| **C. Two-part (FE)** | 1{Duration>0} | YES | YES | -0.001 (0.002) | 0.414 | 1,845,189 |
| | 1{kWh>0} | YES | YES | 0.000 (0.001) | 0.476 | 1,845,189 |
| | ln (Duration) \| Duration>0 | YES | YES | -4.993 (2.152) | 0.020** | 1,646,771 |
| | ln (kWh) \| kWh>0 | YES | YES | -0.325 (3.392) | 0.924 | 106,995 |

### 4.1.2 Amsterdam

Amsterdam exhibits a different pattern. In the two-part specification without fixed effects (Table 6, Panel C: Two-part (No-FE)), a 10% price increase is associated with a substantial reduction in the probability that any charging occurs, while conditional elasticities are smaller in magnitude but precisely estimated.

With station×day and hour-of-week fixed effects (Table 6, Panel C: Two-part (FE)), the extensive-margin association becomes much larger in absolute terms, while conditional intensity responses are not precisely estimated. Taken together, the Amsterdam results indicate that posted prices are most strongly related to the participation decision, while intensity among active hours is comparatively less tightly linked to price once the fixed effects absorb station-day shocks and common timing patterns.

**Table 6. Price–Charging Response (Amsterdam, RQ1)**

| Panel | Outcome | Individual FE | Time FE | Coef (SE) | p | N |
|---|---|---|---|---|---|---|
| **A. No-FE (OLS, HC1)** | Duration (hours) | | | 4.205 (0.064) | 0.000*** | 2,173,376 |
| | Energy (kWh) | | | 60.610 (0.634) | 0.000*** | 2,187,056 |
| | ln (Duration+ε) | | | -0.187 (0.006) | 0.000*** | 2,173,376 |
| | ln (kWh+ε) | | | -0.092 (0.007) | 0.000*** | 2,187,056 |
| **B. FE (station×day + 168h)** | Duration (hours) | YES | YES | -5.125 (9.404) | 0.586 | 2,173,376 |
| | Energy (kWh) | YES | YES | -12.570 (95.331) | 0.895 | 2,187,056 |
| | ln (Duration+ε) | YES | YES | -45.315 (17.881) | 0.011** | 2,173,376 |
| | ln (kWh+ε) | YES | YES | -45.060 (19.582) | 0.021** | 2,187,056 |
| **C. Two-part (No-FE)** | 1{Duration>0} | | | -0.198 (0.001) | 0.000*** | 7,180,920 |
| | 1{kWh>0} | | | -0.198 (0.001) | 0.000*** | 7,180,920 |
| | ln (Duration) \| Duration>0 | | | -0.208 (0.003) | 0.000*** | 3,203,127 |
| | ln (kWh) \| kWh>0 | | | -0.122 (0.003) | 0.000*** | 3,203,237 |
| **C. Two-part (FE)** | 1{Duration>0} | YES | YES | -0.875 (0.249) | 0.000*** | 7,180,920 |
| | 1{kWh>0} | YES | YES | -0.879 (0.249) | 0.000*** | 7,180,920 |
| | ln (Duration) \| Duration>0 | YES | YES | -0.980 (0.974) | 0.314 | 3,203,127 |
| | ln (kWh) \| kWh>0 | YES | YES | -0.958 (0.974) | 0.325 | 3,203,237 |

### 4.1.3 Cross-city synthesis and margin decomposition

Three empirical takeaways emerge from the RQ1 results. First, the extensive margin is more strongly linked to price in Amsterdam than in Shenzhen, especially under the fixed-effects two-part specification, suggesting that participation is the dominant channel of intraday responsiveness in Amsterdam.

Second, Shenzhen's responsiveness, when detectable under fixed effects, is more concentrated in conditional duration, which is consistent with adjustment taking place within active hours rather than through participation.

Third, Figure 7 provides a transparent cross-city summary by decomposing the effect of a common +10% price change on expected kWh into extensive and intensive components, making clear that the dominant margin differs across the two cities.

Overall, RQ1 documents short-run intraday price responsiveness under observed time-of-use regimes and highlights sharp cross-city differences in the margin through which prices are most strongly related to charging behavior. These patterns motivate the subsequent analysis of weather moderation (RQ2) and spatial heterogeneity (RQ3), and they provide inputs for the counterfactual simulations (RQ4), while remaining cautious about fully causal attribution.

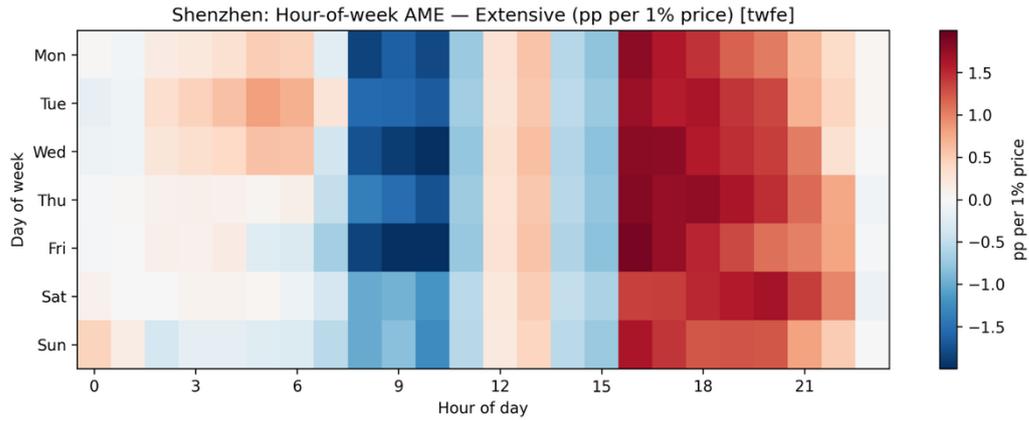

**Figure 3. Shenzhen — Hour-of-week price responsiveness (extensive margin)**

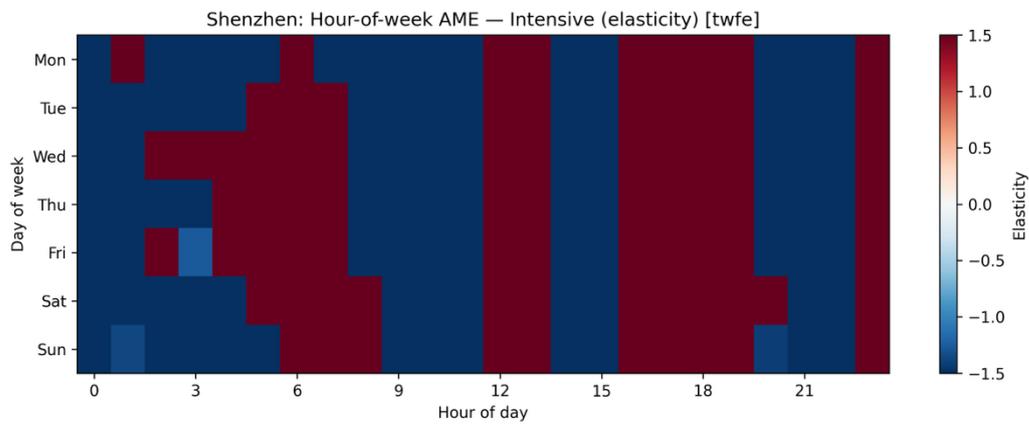

**Figure 4. Shenzhen — Hour-of-week price responsiveness (intensive margin)**

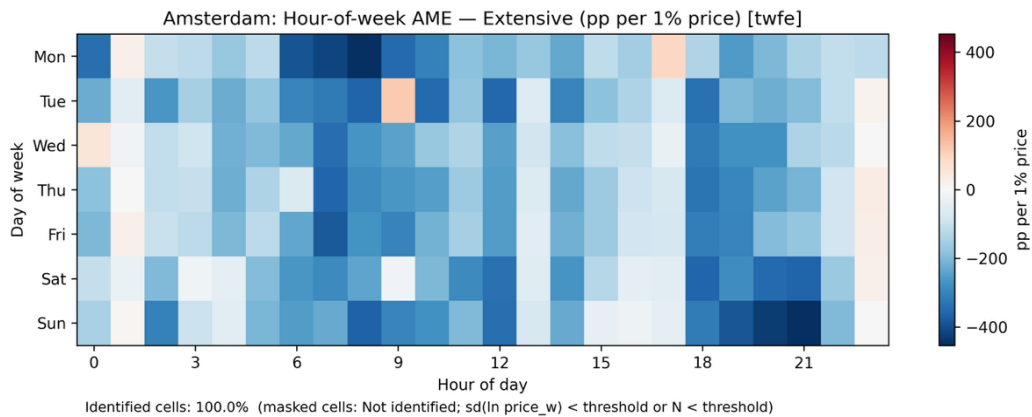

**Figure 5. Amsterdam — Hour-of-week price responsiveness (extensive margin)**

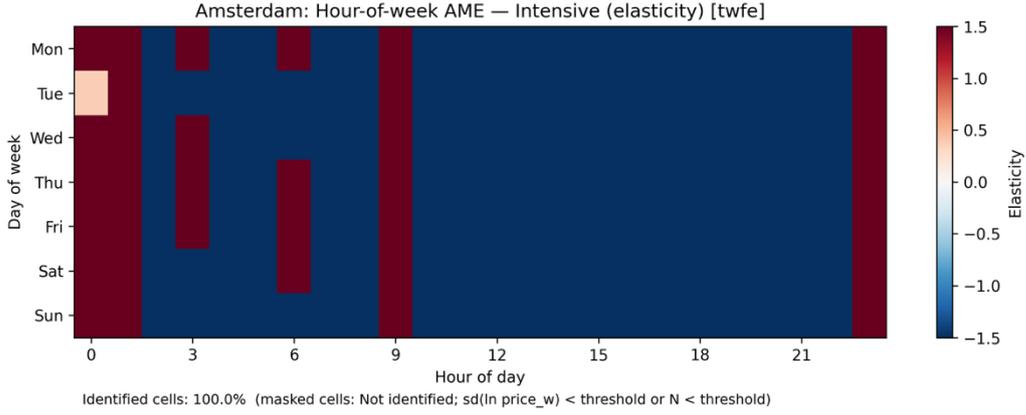

**Figure 6. Amsterdam — Hour-of-week price responsiveness (intensive margin)**

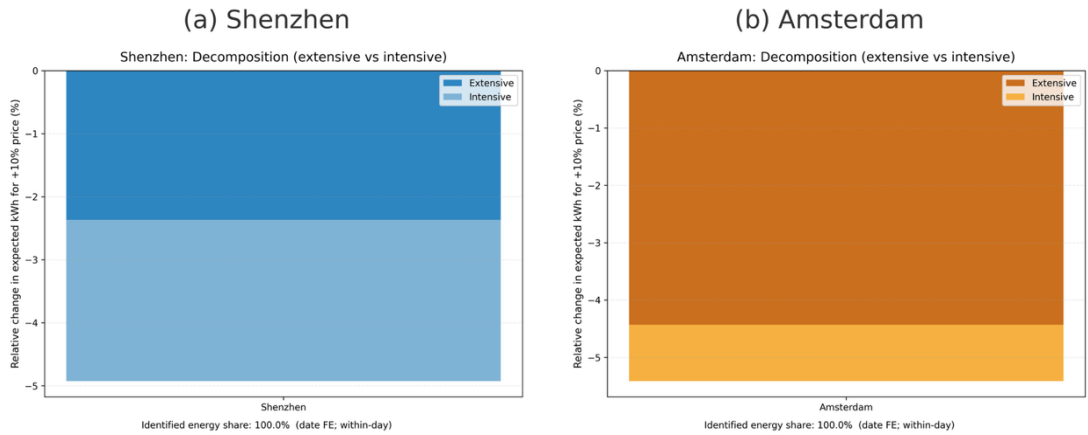

**Figure 7. Decomposition of the price response into extensive and intensive margins.**

### 4.2 RQ2: Weather-dependent heterogeneity in price responsiveness

We examine how temperature and rainfall are associated with heterogeneity in intraday price responsiveness of public charging at both margins, the extensive margin (whether any charging occurs in a station–hour) and the intensive margin (duration or energy conditional on charging). All specifications include station×day and 168 hour-of-week fixed effects, and standard errors are clustered at the station (pile) level. Weather dependence is captured by interacting log price with weather covariates. Throughout, we interpret the interaction patterns as conditional associations in short-run intraday responsiveness under observed time-of-use regimes, rather than as definitive causal moderation effects.

For the intensive margin (log–log conditional models), the effective price slope at hour $h$ can be summarized as the linear combination of the main and interaction terms,

$$\frac{\partial \ln Z}{\partial \ln p}\bigg|_h = \gamma_1 + \gamma_2 T_h + \gamma_3 R_h$$

where $Z \in \{duration, energy\}$, $p$ is the total retail price, and $T_h$, $R_h$ denote temperature and rainfall. A 10% price increase maps to

$$\Delta \ln Z \approx (\gamma_1 + \gamma_2 T + \gamma_3 R) \ln(1.10),$$

evaluated at representative weather states. For the extensive margin (linear probability models), we report the implied change in participation in percentage points using the same linear combination of main and interaction terms multiplied by $ln(1.10)$. Temperature is mean centered within city so that the main price term corresponds to "typical" conditions, and rainfall is coded as an indicator in the baseline (intensity in robustness checks). Coefficients for Shenzhen and Amsterdam are reported in Table 7 and Table 8, respectively. Figure 8 plots implied price elasticities against temperature bins separately for rainy and non-rainy hours, while Figure 9 compares the distributions, showing more negative medians and wider dispersion on rainy hours in both cities.

### 4.2.1 Shenzhen

Shenzhen's warm and humid climate generates frequent hot hours. The estimates indicate that higher temperatures are associated with weaker (less negative) intraday price responsiveness, while rainfall provides a modest and less stable countervailing pattern.

On the extensive margin, higher posted prices are associated with a lower probability that a station–hour records any charging activity. The price×temperature interaction is positive and statistically significant, suggesting that hotter hours attenuate the negative association between prices and participation. As temperature rises above the city mean, the effective price slope becomes flatter. The price×rain interaction is negative and close to conventional significance, consistent with wetter hours being associated with a somewhat more negative effective slope, although precision varies across outcomes.

On the intensive margin, conditional responses move in the same qualitative direction. For duration, the positive and significant price×temperature term indicates that conditional session length is less tightly linked to price when it is hot. Rainfall exerts little incremental association with duration once charging has begun. For delivered energy, the main and interaction effects are less stable in sign and precision under the high fixed-effects design, so interpretation emphasizes the combined slope $\gamma_1 + \gamma_2 T + \gamma_3 R$ evaluated at representative temperatures. At Shenzhen-typical high temperatures, the pattern is a softer conditional response in hot hours, consistent with a larger "must-charge" component that limits short-run flexibility.

Putting these pieces together, hot conditions are associated with weaker estimated price responsiveness on both margins, while rainfall is intermittently associated with more negative effective slopes, especially for the start decision. For peak management in hot climates, TOU signals remain relevant, but the results suggest that non-price complements, such as service guarantees, targeted reservations, or fleet-side coordination, may be needed to maintain reliability without over-penalizing necessary trips.

**Table 7. Weather Moderation of Price Effects (Shenzhen, RQ2)**

| Outcome | Margin | ln(price) | ln(price)×Temp | ln(price)×Rain | N | Individual FE | Time FE |
|---|---|---|---|---|---|---|---|
| Duration | Extensive | −2.829*** | 0.264** | −0.099* | 204,960 | YES | YES |
| kWh | Extensive | −0.000*** | 0.000*** | 0.000*** | 204,960 | YES | YES |
| ln(Duration) | Intensive | 3.116 | 0.611** | 0.107 | 184,123 | YES | YES |
| ln(kWh) | Intensive | 7.587*** | −0.281 | −0.566 | 11,712 | YES | YES |

*Notes.* Extensive = linear probability model for any charging; Intensive = log–log conditional on Y>0. Percent effects for a 10% price step equal coefficient×ln (1.10) (and must be scaled by the baseline participation rate for extensive-margin relative percent interpretations).

### 4.2.2 Amsterdam

Amsterdam's temperate and rain-prone environment exhibits a different pattern from Shenzhen. Inclement weather is associated with stronger (more negative) intraday price responsiveness, most clearly during rainy hours, with additional evidence that warmer conditions also steepen the effective price slope.

On the extensive margin, the price×rain interaction is negative and highly significant, indicating that rainy hours amplify the negative association between prices and participation for a given price increase. The price×temperature interaction is also negative, although less precisely estimated for participation. Taken together, these coefficients imply that the effective participation response to a price change becomes more negative under rain, and, to a lesser extent, at higher temperatures.

On the intensive margin, both interactions are negative and statistically significant in the duration and energy equations. Thus, warmer or wetter hours are associated with larger conditional reductions in charging time and delivered energy when prices rise. Under the high fixed-effects design, this pattern should be read as within-station, within-day heterogeneity in intraday responsiveness, after absorbing station-day shocks and common hour-of-week patterns.

These results suggest a practical operational implication. The posted TOU signal is most tightly linked to behavior precisely when weather raises the generalized cost of charging. In Amsterdam, adverse weather coincides with stronger participation responses and, when charging occurs, sharper conditional reductions, implying greater scope for tariff-based load shaping on inclement-weather hours.

**Table 8. Weather Moderation of Price Effects (Amsterdam, RQ2)**

| Outcome | Margin | ln(price) | ln(price)×Temp | ln(price)×Rain | N | Individual FE | Time FE |
|---|---|---|---|---|---|---|---|
| 1{Duration>0} | Extensive | 1.316 | −0.036 | −0.398*** | 7,180,920 | YES | YES |
| 1{kWh>0} | Extensive | 1.309 | −0.036 | −0.398*** | 7,180,920 | YES | YES |
| ln(Duration) | Intensive | 44.192*** | −1.951*** | −2.546*** | 3,203,127 | YES | YES |
| ln(kWh) | Intensive | 44.632*** | −1.982*** | −2.550*** | 3,203,237 | YES | YES |

*Notes.* Extensive = linear probability model; Intensive = log–log conditional on Y>0. For a 10% price increase, multiply the linear coefficient by ln (1.10) to obtain the change in percentage points (participation) or percent (intensity); for relative-percent participation changes, divide by the baseline participation rate.

### 4.2.3 Cross-city discussion

The two cities exhibit distinct patterns. In Shenzhen, heat is associated with weaker price responsiveness in participation and conditional time, while rain provides at most a modest counterweight. In Amsterdam, rain reliably strengthens the negative association between prices and participation and, together with temperature, deepens conditional responsiveness. Context offers a plausible interpretation for these contrasts. Shenzhen's warm baseline may generate more frequent non-deferrable circumstances that mute short-run adjustments and keep participation responses small, whereas Amsterdam's adverse weather may raise the generalized cost of charging, making prices more likely to tip the decision not to start and to trim intensity once underway. These differences align with the RQ1 evidence that conditional adjustments are more prominent in Shenzhen, while participation plays a larger role in Amsterdam.

### 4.2.4 Robustness and interpretation notes

Results are robust to alternative weather codings (rain as an indicator versus intensity), alternative winsorization thresholds, and restrictions to business hours. Because fixed effects absorb station-day shocks and weekly periodicity, the main price coefficient alone need not align with casual intuition. The economically relevant object is the effective slope $\gamma_1 + \gamma_2 T + \gamma_3 R$ at realized weather states, which we summarize via average marginal effects at representative temperature percentiles and rainy versus non-rainy hours. For transparency, we present both relative effects (log–log) for cross-city comparability and, where informative, absolute changes (levels) for policy translation (e.g., kWh or minutes per +0.10 currency/kWh). Full coefficient tables are reported in Table 7 (Shenzhen) and Table 8 (Amsterdam).

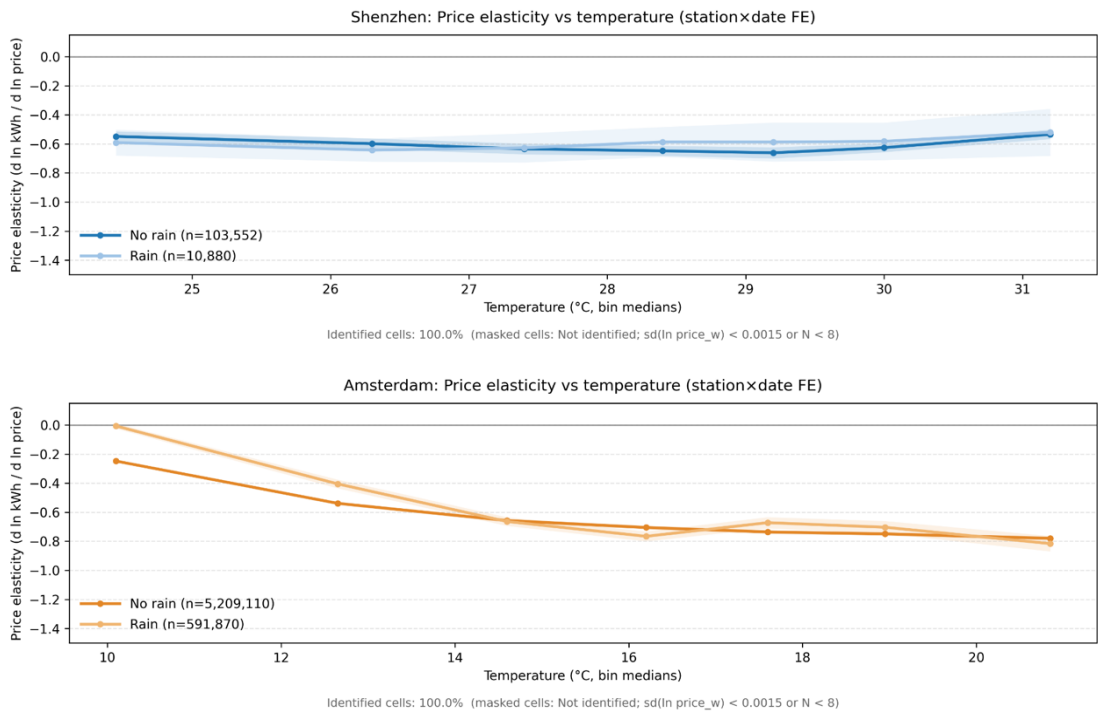

**Figure 8. Price elasticity versus temperature, by precipitation status**

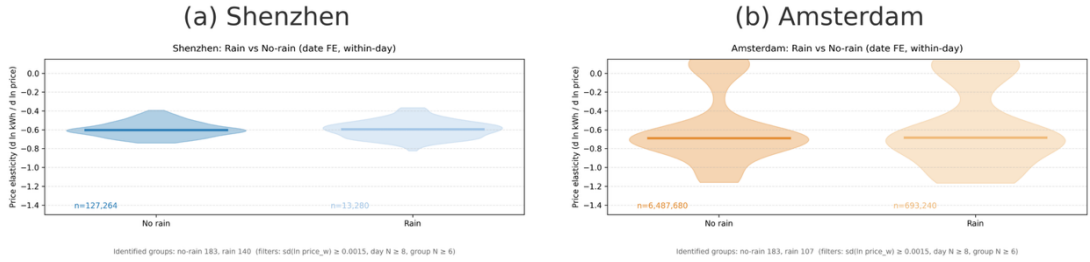

**Figure 9. Price elasticity distributions: rainy vs. non-rainy hours**

## 4.3 Cross-sectional Determinants of Public EV Charging Utilization (RQ3)

We now shift from station–day–hour panels to cross-sectional heterogeneity in realized utilization at the charger level (RQ3). For each city, we aggregate outcomes over the observation window and study per-charger throughput, measured primarily as log total delivered energy (kWh) per charger and, in robustness checks, log total occupied time per charger. The unit of analysis is the individual charger (connector/plug).

We relate per-charger outcomes to three groups of predictors. The first captures equipment capacity, including rated power and on-site scale (the number of chargers at the station). The second captures local network structure, including nearest-neighbor spacing and very-local competition, measured by the number of other public stations within 0.5 km. The third captures the immediate urban context, summarized by the dominant POI background within 500 m. Estimation uses OLS with standard errors clustered at the station level, and coefficients are interpreted as conditional associations rather than causal siting effects, since both equipment and location may respond to latent demand.

Figure 10 maps per-charger energy across the two cities on a common color scale, highlighting pronounced spatial heterogeneity that motivates the regressions below. The specification follows the RQ3 setup in Section 3.5 and keeps variable definitions harmonized across Shenzhen (Section 4.3.1) and Amsterdam (Section 4.3.2). The resulting cross-sectional elasticities and background-specific patterns provide the inputs used in Section 4.4 when simulating counterfactual upgrades and local densification (RQ4).

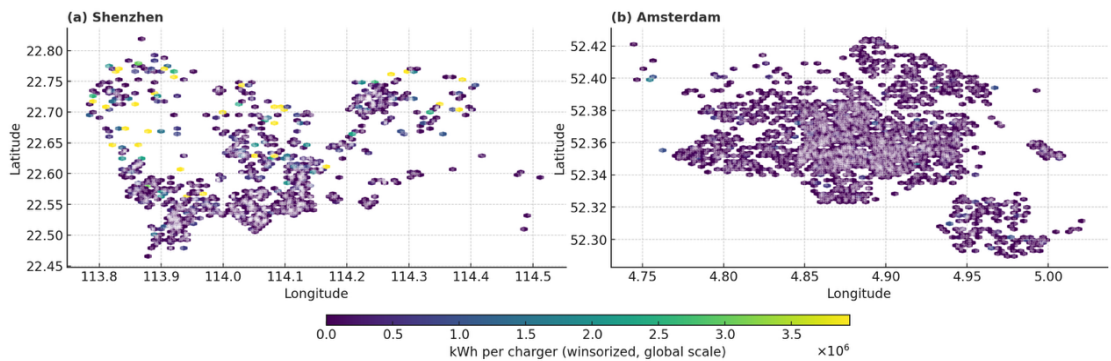

**Figure 10. Spatial distribution of per-charger energy**

### 4.3.1 Shenzhen: Cross-sectional determinants of spatial utilization (RQ3)

We examine cross-sectional differences in realized throughput across Shenzhen's public charging network after accounting for equipment capacity, local network structure, and immediate urban background. The dependent variable is the log of total energy (kWh) delivered per charger over the observation window. Core regressors include log charger power (kW), log station size (number of chargers on site), log nearest-neighbor distance between stations (km), and very-local competition measured as $log\,(1 + number\,of\,other\,public\,stations\,within\,0.5\,km)$. To capture contextual heterogeneity, each station is assigned a dominant POI background within 500 m (commercial, education, transit, health, parking, leisure, civic, or other) based on the largest POI share. Estimation uses OLS with station-clustered standard errors.

Table 9 reports baseline associations at the charger level (N=2,057). The coefficient on $log\,(charger\,power)$ is large and precisely estimated, implying that higher-rated chargers are systematically associated with higher per-charger energy. Very-local competition is negative and statistically significant, indicating that chargers located near many other nearby stations tend to exhibit lower per-charger throughput, conditional on power, size, and broader spacing. Nearest-neighbor distance enters with a negative and marginally significant coefficient, consistent with higher utilization in denser clusters once controls are included, while the station-size coefficient is small and statistically indistinguishable from zero in the baseline model.

To separate where a site is from what a site has, we add POI-background indicators after controlling for charger power, station size, density, and very-local competition. Table 10 reports adjusted utilization levels by background. Relative to commercial locations, education-dominant areas exhibit higher predicted throughput at mean controls, whereas transit-dominant areas exhibit lower predicted throughput; "other" areas are close to commercial once controls are held fixed. For context, Table 11 reports unadjusted distributions by background, highlighting heavy tails and small background sample sizes.

We further allow the very-local competition association to vary by background. Interacting $log\,(1 + competitors \leq 0.5\,km)$ with background dummies (Table 12) reveals sharp heterogeneity: the competition elasticity is strongly negative in education-dominant areas, while it becomes positive in transit-dominant areas. This contrast is consistent with the idea that proximity to other stations is associated with business-stealing in some contexts and with complementary demand-splitting in others, although these patterns should be interpreted as correlational given the endogenous placement of stations and upgrades.

These results should be read as conditional associations rather than causal siting effects, because both location and equipment choices may respond to latent demand. Three design choices reduce mechanical confounding: outcomes are measured per charger, regressions control directly for power and station size, and very-local competition is separated from broader clustering through the nearest-neighbor metric. Robustness checks reported in the Appendix support the qualitative patterns. Finally, the background-specific competition elasticities and adjusted level differences provide the scaling inputs used in Section 4.4 when simulating counterfactual expansion and upgrades.

**Table 9. Shenzhen — Charger-level utilization (RQ3)**

| Variable | (1) Baseline Coef (SE) | (2) + Background FE Coef (SE) |
|---|---|---|
| log(power, kW) | 1.308 (0.056) *** | 1.305 (0.056) *** |
| log(# chargers on site) | 0.030 (0.081) | 0.026 (0.081) |
| log(nearest-neighbor dist., km) | −0.100 (0.051) ** | −0.098 (0.051) ** |
| log(1+competitors ≤0.5 km) | −0.196 (0.055) *** | −0.193 (0.055) *** |
| Education (vs. commercial) |  | 2.035 (0.609) *** |
| Other (vs. commercial) |  | 0.130 (0.269) |
| Transit (vs. commercial) |  | −0.609 (0.383) |
| Observations | 2,057 | 2,057 |
| $R^2$ | 0.373 | 0.379 |

*Notes:* Background FE defined by dominant POI within 500 m (commercial omitted).

*Dependent variable: log (kWh per charger); OLS with station-clustered SEs.*

**Table 10. Shenzhen — Adjusted utilization by POI background (controls at sample means)**

| Background | Predicted mean (kWh/charger) | Diff vs. commercial (%) | N chargers |
|---|---|---|---|
| Commercial | 104,889 | +0.0% | 91 |
| Education | 130,403 | +24.3%* | 13 |
| Other | 103,743 | −1.1% | 1,941 |
| Transit | 76,163 | −27.4% | 11 |

*Notes:* Differences based on background-augmented regression with station-clustered SEs.

*Controls: log(power), log(# chargers), log(nearest-neighbor dist.), log(1+competitors ≤0.5 km).*

**Table 11. Shenzhen — Raw (unadjusted) utilization by POI background (kWh per charger)**

| Background | N | Mean | Median | P25 | P75 |
|---|---|---|---|---|---|
| Commercial | 91 | 1,199,039 | 107,310 | 35,770 | 670,688 |
| Education | 13 | 1,397,101 | 107,310 | 40,880 | 357,700 |
| Other | 1,941 | 902,267 | 71,540 | 35,770 | 214,620 |
| Transit | 11 | 4,622,659 | 498,961 | 95,297 | 4,492,072 |

*Notes:* Unadjusted distributions by dominant POI within 500 m.

**Table 12. Shenzhen — Competition elasticities and background level shifts**

| Background | Competition elasticity | SE | p (comp) | Level shift vs. commercial (log) | SE | p (shift) | N |
|---|---|---|---|---|---|---|---|
| Commercial | −0.003 | 0.323 | 0.990 | 0.000 | 0.000 | 1.000 | 91 |
| Education | −1.275* | 0.417 | 0.000 | 2.035* | 0.609 | 0.001 | 13 |
| Other | −0.194* | 0.050 | 0.000 | 0.130 | 0.269 | 0.633 | 1,941 |
| Transit | +1.443 | 0.607 | 0.017 | −0.609 | 0.383 | 0.110 | 11 |

*Notes:* "Competition elasticity" is the total effect within each background (baseline commercial plus interaction). "Level shift" is the background intercept difference evaluated at zero competition.

*From model with background dummies and background × log(1+competitors ≤0.5 km) interactions; SEs clustered by station.*

### 4.3.2 Amsterdam: Cross-sectional determinants of spatial utilization (RQ3)

We apply the same cross-sectional framework to Amsterdam. The unit of analysis is the individual charger, and the dependent variable is the log of total energy (kWh) delivered per charger over the observation window. Regressors mirror Shenzhen and include log charger power, log station size, log nearest-neighbor distance, and very-local competition measured as $log\,(1+$ $number\ of\ other\ public\ stations\ within\ 0.5\ km)$. To capture context, each station is assigned a dominant POI background within 500 m. In Amsterdam, the baseline category is "other," and the second sizeable category is "parking." Estimation uses OLS with station-clustered standard errors.

The baseline model (Table 13, column 1; N=1,717) yields a large and precisely estimated power elasticity, indicating that higher-rated chargers are systematically associated with higher per-charger energy. Very-local competition is negative and significant, consistent with lower per-charger throughput where many nearby stations are present, conditional on equipment and spacing. Nearest-neighbor distance enters with a negative sign but is not precisely estimated at conventional levels, suggesting at most a mild density premium in Amsterdam once other controls are included. Station size is positive but imprecise.

Adding POI-background indicators (Table 13, column 2) leaves the power and competition coefficients essentially unchanged, suggesting that these associations are not driven by background composition. At mean controls, parking-dominant locations exhibit lower adjusted utilization than "other," although the difference is imprecisely estimated (Table 14). Unadjusted distributions (Table 15) are heavy-tailed with many zeros, underscoring the value of controlling for equipment and local network structure when comparing backgrounds.

Allowing the competition association to vary by background (Table 16) yields negative and statistically significant competition elasticities in both backgrounds, with parking more sensitive than "other." As in Shenzhen, these patterns should be interpreted as conditional associations rather than causal effects of siting, because equipment and location may reflect latent demand and planning decisions. Overall, Amsterdam exhibits a steep power elasticity and consistently negative associations with ultra-local crowding, patterns that we carry forward to Section 4.4 when evaluating counterfactual upgrades and densification.

**Table 13. Amsterdam — Charger-level utilization (RQ3)**

| Variable | (1) Coef (SE) | p | (2) + Background FE Coef (SE) | p |
|---|---|---|---|---|
| log(power, kW) | 1.696 (0.207) *** | 0.000 | 1.685 (0.207) *** | 0.000 |
| log(# chargers on site) | 0.154 (0.102) | 0.131 | 0.160 (0.101) | 0.114 |
| log(nearest-neighbor dist., km) | −0.088 (0.068) | 0.196 | −0.084 (0.068) | 0.218 |
| log(1+competitors ≤0.5 km) | −0.199 (0.060) *** | 0.001 | −0.228 (0.066) *** | 0.001 |
| Parking (vs. other) | | | −0.227 (0.147) | 0.123 |
| Observations | 1,717 | | 1,717 | |
| R² | 0.194 | | 0.197 | |

Dependent variable: log (kWh per charger). OLS; SEs clustered by station. Background baseline = other.

**Table 14. Amsterdam — Adjusted utilization by POI background (controls at sample means)**

| Background | Predicted mean (kWh/charger) | Diff vs. other (%) | p | N chargers |
|---|---|---|---|---|
| Other | 61,054 | +0.0% | (ref.) | 1,563 |
| Parking | 48,649 | −20.3% | 0.123 | 154 |

*Notes:* "Diff vs. other (%)" compares each background to the other baseline; SEs clustered by station.

**Table 15. Amsterdam — Raw (unadjusted) utilization by background (kWh per charger)**

| Background | Mean | Median | N | P25 | P75 |
|---|---|---|---|---|---|
| other | 70,019 | 0 | 3,145 | 0 | 53,830 |
| parking | 49,781 | 0 | 381 | 0 | 35,619 |

*Notes:* Medians of 0 reflect many chargers with no recorded energy in the window; means are heavy tailed.

**Table 16. Amsterdam — Competition elasticities and background level shifts**

| Background | Competition elasticity (SE) | p | Level shift vs. other (log) (SE) | p | N |
|---|---|---|---|---|---|
| Other | −0.202 (0.069) *** | 0.004 | | | 1,563 |
| Parking | −0.444 (0.145) *** | 0.002 | 0.367 (0.428) | 0.390 | 154 |

*Notes:* "Level shift vs. other" is normalized to 0 for the baseline; hence no p-value on the "other" row. Model includes background dummies and background × log(1+competitors ≤0.5 km) interactions; SEs clustered by station; baseline = other.

## 4.4 Policy counterfactuals based on cross-sectional elasticities (RQ4)

We compare equipment upgrades and ultra-local densification using unit changes in expected energy per notional investment. The goal is a practical ranking of options rather than a welfare optimum. These counterfactuals apply the cross-sectional elasticities estimated in RQ3 to observed per-charger baselines, holding network structure and demand composition fixed. They should therefore be read as static projections that help diagnose which levers are likely to be more productive under the observed patterns, not as forecasts of equilibrium outcomes.

We project system impacts by mechanically mapping each scenario into a change in per-charger baseline energy. Let $Y_i$ denote baseline energy per charger. A proportional power upgrade of size $\Delta P$ (e.g., $\Delta P = 0.10$ for +10%) maps to

$$Y_i^{cf} = Y_i \cdot (1 + \varepsilon_p \Delta P)$$

where $\varepsilon_p$ is the power elasticity estimated in RQ3.

For ultra-local densification, the relevant regressor is $log(1+c)$, where ccc is the number of other public stations within 0.5 km. Adding one nearby competitor changes the regressor from $log(1+c)$ to $log(1+c+1) = log(c+2)$, so the implied increment is

$$\Delta log(1+c) = log(c+2) - log(c+1) = log\left(\frac{c+2}{c+1}\right).$$

The projected per-charger energy under a +1 competitor shock is therefore

$$Y_i^{cf} = Y_i \cdot exp(\varepsilon_{comp}^g \cdot \Delta log(1+c)),$$

where $\varepsilon_{comp}^g$ is the background-specific competition elasticity (with $g$ indexing the dominant POI background within 500 m). We apply scenario definitions consistently across cities (Tables 17 and 20), report citywide totals (Tables 18 and 21), and summarize by-background changes (Tables 19 and 22).

### 4.4.1 Shenzhen

Shenzhen exhibits a large and precisely estimated power elasticity ( $\varepsilon_p \approx 1.308$ ) and highly heterogeneous associations for ultra-local densification: the competition elasticity is positive in transit-dominant micro-environments but strongly negative in education-dominant areas. We evaluate five scenarios: S1 (+10% power for all chargers), S2 (+50% power for bottom-quartile chargers), S3 (+1 very-near competitor in transit), S4 (+1 very-near competitor in education), and S5 (S2+S3).

Citywide projections are reported in Table 18. A uniform +10% power upgrade (S1) is projected to increase total delivered energy by +13.1%. Targeting upgrades to the bottom power quartile (S2) yields a +2.5% increase, indicating that concentrating upgrades on low-power devices can still deliver meaningful gains under budget constraints.

Projected effects of ultra-local densification depend strongly on background (Tables 18–19). Adding one nearby station in transit-dominant areas (S3) increases city totals by +4.7% and substantially raises throughput within the transit background (+147.2%). In contrast, adding one nearby station in education-dominant areas (S4) slightly reduces city totals (−0.5%) and lowers education-background throughput (−41.4%). The combined scenario (S5) yields a +7.2% citywide increase, reflecting that projected upgrade gains are broad-based, while densification benefits are concentrated in specific contexts.

These patterns motivate an operational interpretation that remains consistent with the associational nature of the inputs. In Shenzhen, upgrades appear to be a robust lever across backgrounds, while densification is projected to be beneficial primarily in transit-like corridors and potentially counterproductive in education-like micro-environments.

**Table 17. Shenzhen — RQ4 scenario definitions**

| Scenario | What changes? | Implementation (elasticities used) |
|---|---|---|
| S1 | +10% power for all chargers | $Y_i \times (1 + 1.308 \times 0.10)$ |
| S2 | +50% power for bottom-quartile chargers | If power ≤ Q1: $Y_i \times (1 + 1.308 \times 0.50)$ |
| S3 | +1 competitor (transit only) | Transit: $Y_i \times exp(1.443 \cdot \Delta log(1+c))$ |
| S4 | +1 competitor (education only) | Education: $Y_i \times exp(-1.275 \cdot \Delta log(1+c))$ |
| S5 | S2 + S3 | Apply S2 then S3 multiplicatively |

**Table 18. Shenzhen — Citywide outcomes (kWh; Δ vs baseline)**

| Scenario | Total kWh (baseline) | Total kWh (counterfactual) | Δ total (%) |
|---|---|---|---|
| S1: +10% power (all) | 1,929,459,272 | 2,181,832,545 | +13.1% |
| S2: +50% power (bottom quartile) | 1,929,459,272 | 1,978,169,446 | +2.5% |
| S3: +1 competitor (transit only) | 1,929,459,272 | 2,020,471,648 | +4.7% |
| S4: +1 competitor (education only) | 1,929,459,272 | 1,919,318,324 | −0.5% |
| S5: S2 + S3 | 1,929,459,272 | 2,069,227,599 | +7.2% |

**Table 19. Shenzhen — By-background percentage changes (Δ% vs background baseline)**

| Scenario \ Background | commercial | education | other | transit |
|---|---|---|---|---|
| S1 | +13.1% | +13.1% | +13.1% | +13.1% |
| S2 | +3.5% | +0.4% | +2.6% | +0.1% |
| S3 | +0.0% | +0.0% | +0.0% | +147.2% |
| S4 | +0.0% | −41.4% | +0.0% | +0.0% |
| S5 | +3.5% | +0.4% | +2.6% | +147.3% |

*Notes:* Effects are static projections from RQ3 elasticities; background taxonomy corresponds to §4.3.1.

### 4.4.2 Amsterdam

Amsterdam features an even steeper estimated power elasticity ($\varepsilon_p \approx 1.685$) and negative ultra-local competition elasticities across backgrounds, with parking-dominant areas more sensitive than the "other" baseline. We mirror the five scenarios, now targeting densification to parking (S3) and other (S4). Because power has a mass point at 6.4 kW, we define the "bottom quartile" in S2 as the rank-based bottom 25% among chargers with positive power to avoid mechanical sensitivity to the mass point.

Citywide projections (Table 21) show that a uniform +10% power upgrade (S1) increases total delivered energy by +16.8%, larger than Shenzhen in line with the higher $\varepsilon_p$. Targeted upgrades (S2) yield a +3.6% increase, with larger gains in parking than in other (Table 22), consistent with low-power devices in parking contexts being more upgrade-responsive under the estimated elasticities.

Ultra-local densification is projected to reduce totals in Amsterdam. Adding one nearby station in parking areas (S3) lowers city totals by −0.6% and reduces parking-background throughput by −7.4%. Densifying the "other" background (S4) yields a −1.2% citywide change. The combined scenario (S5) nets +3.0%, implying that densification offsets part of the upgrade benefit in parking micro-environments under the static projection.

Taken together, the Amsterdam projections suggest that upgrades dominate densification as a lever in this setting, while face-to-face siting in parking-dominant corridors is less attractive unless additional evidence indicates unmet demand or congestion that is not captured by the cross-sectional associations.

**Table 20. Amsterdam — RQ4 scenario definitions**

| Scenario | Implementation |
|---|---|
| S1: +10% power (all chargers) | $Y_i \times (1 + 1.685 \times 0.10)$ |
| S2: +50% power (bottom quartile only) | Bottom 25% (rank among positive − power) $\times (1 + 1.685 \times 0.50)$ |
| S3: +1 competitor (parking only) | Parking: $exp(-0.444 \cdot \Delta log(1+c))$ |
| S4: +1 competitor (other only) | Other: $exp(-0.202 \cdot \Delta log(1+c))$ |
| S5: S2 + S3 combined | Apply S2 then S3 multiplicatively |

**Table 21. Amsterdam — Citywide outcomes (kWh; Δ vs baseline)**

| Scenario | Total kWh (baseline) | Total kWh (counterfactual) | Δ total (%) |
|---|---|---|---|
| S1: +10% power (all chargers) | 239,175,391 | 279,476,445 | +16.8% |
| S2: +50% power (bottom quartile only) | 239,175,391 | 247,813,521 | +3.6% |
| S3: +1 competitor (parking only) | 239,175,391 | 237,774,624 | −0.6% |
| S4: +1 competitor (other only) | 239,175,391 | 236,390,117 | −1.2% |
| S5: S2 + S3 | 239,175,391 | 246,306,776 | +3.0% |

**Table 22. Amsterdam — By-background outcomes**

| Scenario | Background | Baseline kWh | Counterfactual kWh | Δ (%) |
|---|---|---|---|---|
| S1 | other | 220,208,709 | 257,313,876 | +16.9% |
|    | parking | 18,966,683 | 22,162,569 | +16.9% |
| S2 | other | 220,208,709 | 227,658,534 | +3.4% |
|    | parking | 18,966,683 | 20,154,986 | +6.3% |
| S3 | other | 220,208,709 | 220,208,709 | +0.0% |
|    | parking | 18,966,683 | 17,565,916 | −7.4% |
| S4 | other | 220,208,709 | 217,423,434 | −1.3% |
|    | parking | 18,966,683 | 18,966,683 | +0.0% |
| S5 | other | 220,208,709 | 227,658,534 | +3.4% |
|    | parking | 18,966,683 | 18,648,242 | −1.7% |

*Notes:* Elasticities from §4.3.2: power =1.685; competition: other = −0.202, parking = −0.444. S2 uses rank-based bottom quartile among positive-power chargers to avoid mass-point bias at 6.4 kW.

### 4.4.3 Cross-city synthesis and policy implications

Two robust patterns emerge across cities. First, equipment upgrades are consistently projected to increase throughput: a uniform +10% power upgrade raises totals by +13.1% in Shenzhen and +16.8% in Amsterdam, while targeted upgrades to low-power devices still deliver +2.5% and +3.6%, respectively. This cross-setting consistency reflects the large estimated power elasticities in both networks.

Second, ultra-local densification is highly context dependent. In Shenzhen, densifying transit corridors is projected to be complementary, while densifying education-dominant areas appears cannibalizing. In Amsterdam, densification is projected to be negative across backgrounds, especially in parking-dominant areas. These contrasts align with the background-specific competition elasticities estimated in RQ3 (positive in Shenzhen-transit, negative elsewhere and across Amsterdam).

A practical operational reading is therefore to treat upgrades as the first-line lever and to apply a context gate for densification. Under limited budgets, prioritizing low-power upgrades offers measurable gains, and selective infill is most defensible where background-specific estimates and observable congestion indicators jointly suggest complementarity rather than substitution. Finally, these counterfactuals are static and hold network and demand structure fixed. They may understate upgrade benefits if higher power reduces queues that unlock time-shifted demand, and they may overstate densification gains in substitution-heavy contexts. Robustness checks (median-based outcomes and alternative winsorization) preserve the qualitative ranking.

## 4.5 Robustness Check

This section assesses whether our main findings are sensitive to alternative ways of measuring prices and outcomes, to the granularity of fixed effects, to trimming and functional-form choices, and to sample restrictions used for RQ1–RQ4. Throughout, these exercises are designed to evaluate stability and internal consistency of the estimated short-run intraday price responsiveness and the cross-sectional patterns used for RQ3–RQ4. They do not convert the analysis into a fully causal design, but they help clarify which components of variation are driving the estimates and whether qualitative conclusions persist under reasonable perturbations.

### 4.5.1 Identification and the availability of within-cell price variation (RQ1–RQ2)

A central requirement of our RQ1–RQ2 design is that estimated price responsiveness is driven by within fixed-effect cells, rather than by cross-site or cross-day contrasts. We therefore audit the variance of posted prices after absorbing site×date fixed effects. Concretely, we decompose the variance of posted price into the component explained by site×date fixed effects and the residual component that remains within site×date cells.

In both cities, site×date fixed effects explain essentially none of the posted-price variance, leaving nearly all variation in the residual. Figures 11 (Shenzhen) and 12 (Amsterdam) illustrate this pattern: the residual share approaches 100%. This indicates that, even after conditioning on site×date fixed effects, there remains ample within-cell price movement across hours (and, where applicable, across connectors or tariff tiers). The audit supports the empirical feasibility of identifying short-run intraday price responsiveness from within-cell variation under the fixed-effect structure used in RQ1–RQ2.

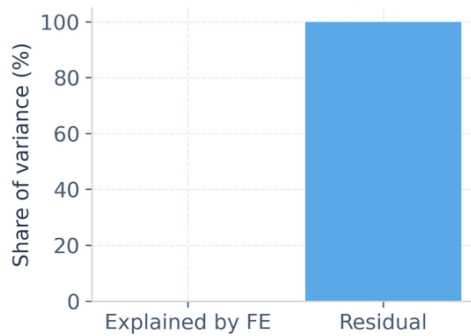

**Figure 11. Shenzhen—price-variation audit under site×date FEs**

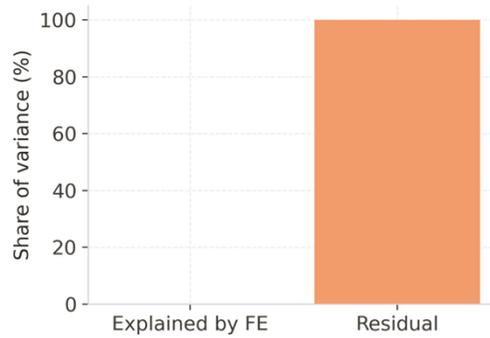

**Figure 12. Amsterdam—price-variation audit under site×date Fes**

### 4.5.2 Consistency of price elasticities across specifications and subsamples (RQ1–RQ2)

We next examine whether estimated elasticities remain stable across a broad grid of modeling choices. We re-estimate the main price–charging relationships under alternative price measures (posted vs. energy-normalized), alternative fixed-effect structures (e.g., station×day with hour-of-week vs. station×week with hour), alternative temporal handling (hourly vs. 2-hour aggregation; with and without one-hour and one-day lags), alternative functional forms (log–log vs. semi-log), alternative trimming rules (1–99% vs. 0.5–99.5%), and alternative weather controls and interaction specifications (temperature terciles and spline variants).

Two additional falsification-style checks are included to probe whether the results hinge mechanically on time structure rather than within-cell price movements. First, we implement calendar-shift permutations that reshuffle hours within stations. Second, we run rolling pre-trend checks in which the same estimation is applied to placebo timing windows. Across these exercises, the placebo estimates center near zero, consistent with the view that the identifying signal in the main specifications is tied to within-cell intraday price variation rather than arbitrary hour patterns.

Across the full set of alternatives, the qualitative conclusions remain unchanged. Estimated elasticities are negative in both cities, and peak versus off-peak estimates are statistically close with overlapping 95% confidence intervals. Price responsiveness is strongest under low-temperature conditions, and Amsterdam generally exhibits larger magnitudes (in absolute value) than Shenzhen. The overall and subgroup estimates with 95% confidence intervals are summarized in Figures 13–14, while the temperature-tercile patterns are shown in Figures 15–16. Numerical subgroup estimates for Amsterdam are reported in Table 23.

**Table 23. Amsterdam—Price elasticity by subgroup (main specification)**

| Group | Elasticity (d ln(1+kWh)/d ln price) | SE | 95% CI low | 95% CI high |
|---|---|---|---|---|
| Overall | −0.569 | 0.225 | −1.011 | −0.127 |
| Peak | −0.496 | 0.218 | −0.923 | −0.069 |
| Off-peak | −0.612 | 0.232 | −1.067 | −0.156 |
| Low T | −0.636 | 0.235 | −1.097 | −0.176 |
| Mid T | −0.569 | 0.222 | −1.004 | −0.133 |
| High T | −0.444 | 0.196 | −0.828 | −0.061 |

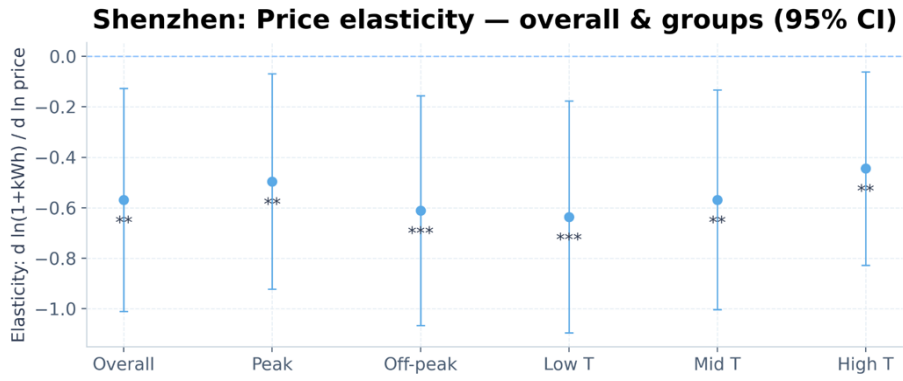

Figure 13. Shenzhen—forest plot: overall & subgroup elasticities, 95% Cis

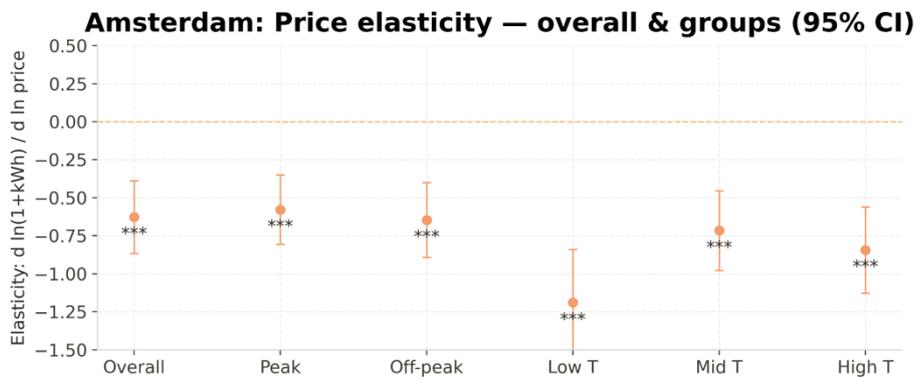

Figure 14. Amsterdam—forest plot: overall & subgroup elasticities, 95% Cis

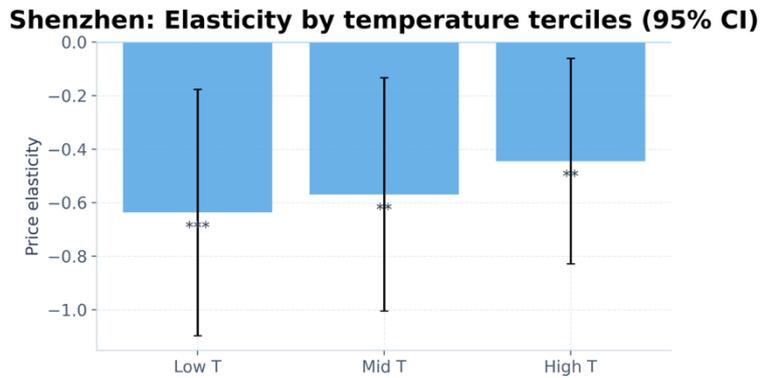

Figure 15. Shenzhen—elasticities by temperature terciles, 95% Cis

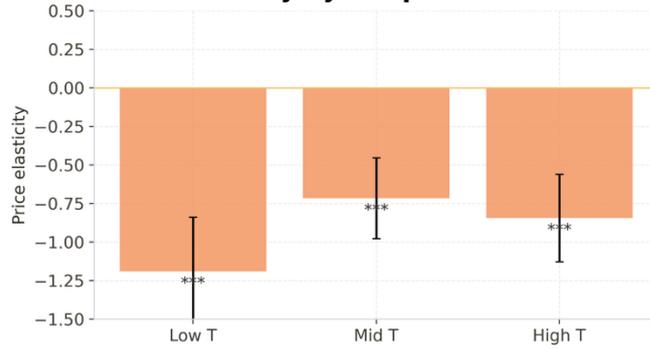

Figure 16. Amsterdam—elasticities by temperature terciles, 95% Cis

**4.5.3 Spatial heterogeneity and network accessibility in cross-section (RQ3)**

For RQ3, we test whether the cross-sectional associations between utilization and local network structure are sensitive to alternative definitions of neighborhood exposure and outcome scaling. Using station-level cross sections, we regress $log\,(1 + kWh)$ on $log\,effective\,capacity\,(ports \times average\,port\,power)$, $log(1 + nearby - station\,counts)$ within a local radius, and land-use indicators, with robust (HC1) standard errors.

Varying the neighborhood radius between 0.5 and 1.0 km, switching trimming between 1–99% and 0.5–99.5%, and replacing the outcome with $log\,(1 + kWh/port)$ leaves the qualitative patterns intact. Denser local networks remain negatively associated with own-station throughput, consistent with spatial substitution at very local scales, while effective capacity is strongly and positively associated with use. The core 0.5 km specification is summarized in Table 24. These cross-sectional checks reinforce that the RQ3 patterns are not driven by a single radius choice or by a particular outcome scaling, while remaining interpretable as conditional associations rather than causal siting effects.

**Table 24. RQ3 core cross-section (radius 0.5 km; trimming 1–99%)**

| City | N | β_neighbors (SE) | β_effective capacity (SE) | R² |
|---|---|---|---|---|
| Shenzhen | 1348 | −0.095 (0.044) | 1.262 (0.044) | 0.360 |
| Amsterdam | 1341 | −0.130 (0.041) | 1.741 (0.086) | 0.278 |

**4.5.4 Counterfactual network augmentation and stability of scenario rankings (RQ4)**

Finally, we verify that the RQ4 scenario rankings are not an artifact of a particular capacity definition or neighborhood radius. We recompute citywide utilization changes under the five scenarios using alternative capacity definitions (nameplate vs. derated power) and alternative neighbor radii (0.5–1.0 km) and confirm that qualitative rankings and within-category signs remain stable.

For Shenzhen, the citywide percentage changes are preserved when switching these implementation choices, and the ordering of scenarios remains unchanged (Table 25 and Figure 17). In addition, within-category patterns retain their signs, with large positive changes concentrated in transit contexts and negative changes concentrated in education contexts. For Amsterdam, applying the same scenario structure yields the same qualitative message: upgrades generate sizeable projected gains, while the densification scenarios remain mechanically near zero under the current POI mapping because no stations are tagged as transit/education in Amsterdam in this classification; in that case, S5 mirrors the corresponding upgrade component. The compiled percentage changes are reported in Table 25, with bar-chart summaries in Figures 17–18.

Taken together, the robustness exercises support the internal consistency of the main empirical patterns for RQ1–RQ4 under reasonable variations in measurement and specification. They do not establish a fully causal design, but they provide evidence that the key qualitative conclusions are not driven by a narrow set of modeling choices.

**Table 25. RQ4—Citywide counterfactual impacts (% change in total kWh)**

| City / Scenario | S1: +10% power | S2: +50% bottom-25% | S3: +1 neighbor (transit) | S4: +1 neighbor (education) | S5: S2+S3 |
|---|---|---|---|---|---|
| Shenzhen | +13.1% | +2.5% | +4.7% | −0.5% | +7.2% |
| Amsterdam | +16.2% | +17.9% | 0.0% | 0.0% | +17.9% |

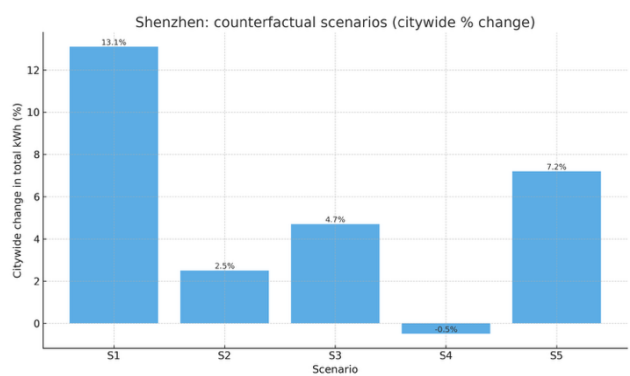

Figure 17. Shenzhen—RQ4 citywide impacts

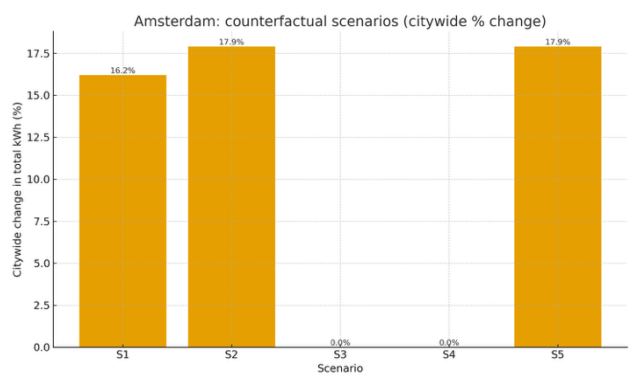

Figure 18. Amsterdam—RQ4 citywide impacts

# 5 Discussion

## 5.1 Economic interpretation and identification scope

This study describes how public EV charging outcomes move with posted time-of-use prices within the day. The main object is short-run adjustment. It is about retiming and reallocation across nearby hours, rather than long-run changes such as EV adoption, home charging access, or location choice. The estimates are therefore most useful for day-to-day operations, peak management, and near-term planning.

Our framework separates two behavioral margins. The extensive margin asks whether any charging happens in a station hour. The intensive margin asks how much energy or time is used once charging starts. This split matters because similar average elasticities can hide different channels. In some settings, pricing mainly affects whether drivers start a session. In other settings, charging is more necessary, and pricing mainly affects how long users stay or how much energy they take.

Identification uses within station-by-day price movements induced by scheduled time-of-use regimes, together with fixed effects. Station-by-day fixed effects absorb station-specific conditions that are stable within the same day, such as local demand level, outages that last the day, and other daily operating conditions. Hour-of-week fixed effects absorb predictable weekly rhythms in charging. Under this design, the estimated price slope reflects how charging outcomes move with price changes within the same station and day, after removing stable daily factors and regular weekly patterns.

At the same time, we do not treat these patterns as clean proof of causality. Even with rich fixed effects, some unobserved station-hour factors can still vary within a day and move together with price-relevant conditions. In addition, users' choice sets and search costs are not directly observed. For these reasons, we interpret the results as strong, systematic associations that are consistent with short-run behavioral response under scheduled pricing, rather than as a definitive causal parameter that would hold under any pricing system.

Even with rich fixed effects, our estimates are best interpreted as short-run responses of public charging outcomes to posted prices within the observed tariff regimes. What we do not identify is the mechanism behind those responses. When charging falls after a price increase, it may reflect shifting to other public stations, switching to home or workplace charging, delaying charging to another hour, or reducing total charging demand. Weather interactions show that the size of the response changes across conditions, but the analysis cannot tell which specific behavioral channel drives that change.

**5.2 Mechanisms behind heterogeneity across cities, weather states, and space**

Across both cities, charging outcomes move with posted prices in economically meaningful ways. The two-part results show that adjustment can come from starting behavior, from within-session behavior, or from both. This helps explain why one city can look more "price responsive" on average while the underlying channel is different.

The weather results add another layer. In Shenzhen, hotter conditions are linked to weaker price responsiveness. A simple interpretation is that high heat increases the "must charge" component. When charging becomes more necessary, users have less room to shift or cut usage in the short run. Rain in Shenzhen provides at most a limited reinforcement, and the patterns are less stable than temperature effects.

In Amsterdam, inclement weather is linked to stronger price responsiveness, especially in rainy hours. A simple interpretation is that bad weather raises the generalized cost of charging trips. When the outside conditions are worse, a given price increase is more likely to tip the decision away from starting, and it can also reduce intensity once charging occurs. This contrast is consistent with the idea that the same price schedule can operate differently when the local environment changes the ease of substitution.

The cross-sectional results in RQ3 help connect these patterns to network structure. Higher charger power is strongly associated with higher realized throughput per charger in both cities. Nearby competition tends to be associated with lower throughput, but the strength and even the sign can depend on context. This is important for interpretation. Dense networks can lower search costs and make substitution easier, but they can also split demand across close sites. Because siting and equipment choices respond to latent demand, we treat these relationships as conditional associations, not causal effects of infrastructure.

Finally, the counterfactual scenarios in RQ4 should be read in the same spirit. They are mechanical projections that apply the estimated cross-sectional elasticities to rank options, rather than forecasts of what would happen under a full equilibrium response. Even under this cautious reading, one pattern is very stable. Upgrading power at existing sites tends to produce larger gains than adding very nearby sites, and densification outcomes depend strongly on where it is done.

**5.3 From estimates to operational levers and policy design**

Within the scope of observed time-of-use regimes and short-run behavior, the findings point to practical levers. These are not prescriptions for a long-run optimal tariff. They are simple lessons for operations and near-term planning.

First, treat time-of-use pricing as a tool for shifting load across hours, not as a guarantee of large demand reduction. The estimates show that charging activity can move with price, but the amount of movement depends on local flexibility. Where users are captive or constraints bind, stronger differentials may mainly increase bills without moving much load. In those settings, service reliability and capacity become more important than further price steepening.

Second, prioritize upgrades when power or congestion is a binding constraint. The cross-sectional evidence and the RQ4 scenarios both point to high returns from increasing effective power at existing sites. This lever is also easy to target. It does not rely on users switching locations. It mainly raises the ceiling on what the system can deliver.

Third, be careful with densification at very short distances. Adding nearby stations can help in some contexts, but it can also dilute demand and reduce throughput per charger. The Shenzhen results suggest that the same "add one nearby site" action can look very different across micro-environments. In Amsterdam, the evidence is more consistently negative for very local crowding, especially in parking-dominant areas. A practical takeaway is to gate densification by context and by observed local conditions. Densification is more defensible where there is clear unmet demand or persistent congestion, and where users can easily substitute across nearby options.

Fourth, incorporate weather into operational planning. The weather interaction results indicate that price responsiveness can change with temperature and rain. This matters for reliability. In hot climates, the short-run ability to shift charging may be limited exactly when the system is stressed. In rainy or bad-weather conditions, some users may become more sensitive to prices and less willing to start a session. Operators can use this to plan staffing, maintenance windows, and communication, and to avoid applying strong price pressure in conditions where users have low flexibility.

Fifth, keep equity and user burden in view. Uniform schedules can impose higher costs on users who cannot shift easily because of where they live, when they travel, or how exposed they are to weather. Because our analysis focuses on short-run responses, it does not deliver a full welfare evaluation. Still, the heterogeneity patterns suggest a practical principle. Apply stronger price differentials where flexibility is higher, and substitution is easier. Use reliability measures and targeted upgrades where flexibility is low. If tariff reforms aim to increase peak to off-peak gaps, revenue-neutral designs that stabilize average daily bills can reduce unnecessary burden.

Overall, the main contribution of the paper is to show, at station-hour resolution and in two very different cities, how charging outcomes move with posted prices, and how this link depends on weather and place. The results help explain why a single "average elasticity" can be misleading for operations. They also show why infrastructure, and pricing should be designed together, since pricing works best when non-price constraints are not binding.

# 6 Conclusions

This paper studies public EV charging under time of use pricing in two cities, Shenzhen and Amsterdam, using station by day by hour data. The results show that charging outcomes respond to posted prices within the day in systematic ways. The magnitudes are most relevant for short run operations under the observed tariff regimes and network conditions.

A central message is that the same price schedule can work through different behavioral margins. In Shenzhen, price variation is linked mainly to what happens after charging begins. Delivered energy and charger occupancy change with price more than the start decision. In Amsterdam, the price link is strongest on the participation margin. Higher prices are associated with fewer station hours with any charging, while conditional intensity changes less once charging occurs. This difference matters for practice because it changes what a tariff can realistically achieve. In a participation driven system, prices mainly shape when and where sessions start. In an intensity driven system, prices mainly shape how sessions use time and energy once they start.

Responsiveness is not constant across conditions. Peak hours in both cities show weaker responses, consistent with tighter constraints when the system is busy. Weather shifts the strength of the response in different directions across the two environments. Hotter hours in Shenzhen coincide with weaker price responsiveness. Rainy hours in Amsterdam coincide with stronger participation responses. These patterns suggest that the operational leverage of time of use pricing depends on conditions that change short run flexibility.

Where stations are located also matters. Commercial and transit-oriented areas generally show stronger movement of charging outcomes with price than residential or education-oriented areas. The form of spatial variation differs across cities. Shenzhen's spatial heterogeneity shows up mainly in intensity responses, while Amsterdam's shows up mainly in participation responses. This points to a role for land use context and local network structure in shaping users' ability to adjust timing and location within the day.

Cross sectional utilization results provide additional guidance for near term planning. Higher charger power is strongly linked to higher realized throughput per charger in both cities. Very local crowding is often linked to lower throughput, with meaningful variation by local context. Scenario comparisons that apply these cross-sectional elasticities to observed baselines suggest that targeted upgrades at existing sites deliver more consistent gains than very local densification on average, while the effect of adding nearby sites can be positive in some contexts and negative in others. These scenario results are best viewed as structured comparisons under current patterns, not as forecasts.

Finally, the analysis clarifies what is measured and what remains unknown. We quantify how public charging outcomes respond when posted prices change in the short run. We do not observe the channel behind the response. A decline in public charging after a price increase could come from shifting to other public stations, switching to home or workplace charging, delaying charging to another hour, or reducing total charging demand. The weather patterns may operate through any of these channels. Data that link station outcomes to user choices, alternative charging options, and queueing conditions would be needed to separate mechanisms and to evaluate welfare and equity impacts of pricing and infrastructure design.

# 7 Limitations and Outlook

This study has several limitations that matter for interpretation, especially when we discuss mechanisms. The analysis uses high frequency station by day by hour outcomes in two cities and focuses on short run, within day responses under existing public charging networks and observed time of use price schedules. The estimates describe how charging outcomes move when posted prices change within the same station and day, after controlling for stable station differences and predictable weekly patterns. They are most useful for near term operations, such as load shifting and congestion management. They do not speak to longer run adjustments, including EV adoption, expansion of home or workplace charging, or changes in routine travel and location choice.

A first limitation concerns what we can and cannot say about mechanisms. With the fixed effect design, we can interpret the estimates as the effect of posted price changes on observed station hour charging outcomes in the short run, within the settings covered by our data and pricing regimes. However, the analysis cannot explain *why* those outcomes change. When energy or occupancy falls after a price increase, we cannot tell whether users reduce total charging, shift to other hours, switch to nearby public stations, or substitute to other options such as home or workplace charging. Similarly, when weather changes the price responsiveness, we can document that the relationship differs across weather states, but we cannot identify which behavioral pathway drives the difference. The results therefore describe causal responses in outcomes to price changes in the observed public networks, but they remain reduced form with respect to the underlying decision process.

A second limitation is that charging behavior is observed only through delivered energy and occupancy time. These measures are directly relevant for network operations, but they provide an incomplete view of user decisions. The data do not observe who is charging, what the charging purpose is, what trip context they face, or what alternatives they have. Prior work suggests that public charging choices depend on user characteristics, attitudes, trip needs, and station attributes, and that these determinants can vary across settings (Potoglou et al., 2023). Reviews of EV charging behavior also stress strong heterogeneity in preferences and constraints across users and situations, which is difficult to recover from station level outcomes alone (Shariatazadeh et al., 2025).

This measurement issue is central for mechanisms because each station hour aggregates decisions by different user types and charging purposes. Public charging can include drivers without home charging, users topping up during activities, and commercial or fleet users who need reliable within day charging. Changes in the mix of these users can affect both the participation margin and the conditional intensity margin. The user mix can also differ across cities, land use contexts, and time of day. Related work highlights that reliance on public charging is linked to housing type and access to home charging, so dependence on public networks can vary sharply across groups (Kristoffersson et al., 2025). This is also a key concern in recent equity research on public charging coverage and reliability, which emphasizes that limited access to home charging can increase dependence on public networks and widen disparities (Yu et al., 2025).

A third limitation is that the empirical framework does not explicitly model queueing, waiting, and strategic interaction across users. Station by day fixed effects reduce concerns about many short run shocks, but they do not capture how real time congestion information or expectations about availability shape decisions to start charging, to leave, or to search for alternatives. Queueing behavior can generate patterns that look like price sensitivity in participation even when the main driver is congestion avoidance. Queueing models in charging settings highlight behaviors such as balking and reneging, which can directly affect observed participation and occupancy outcomes (Varshney et al., 2025). Without direct measures of waiting time, queue length, or app-based availability information, these channels remain untested.

A fourth limitation is that weather enters as an observed state variable, but the analysis does not observe the pathways through which weather matters. Weather may change travel demand and timing. It may also change walking costs, search costs, and the perceived value of time. It can also trigger anticipatory behavior, such as charging earlier when bad conditions are expected. Distinguishing these explanations would require data on trip purpose, time constraints, expectations, and real time station conditions, which are not available here.

These limitations point to several concrete directions for future research. One direction is to link station outcomes to user identity and charging purpose, for example by using anonymized vehicle identifiers, membership classes, fleet flags, or proxies for home charging access. This would allow separation of responses across private and commercial users, and across users with and without viable alternatives. A second direction is to integrate operations data that measure congestion and availability, such as occupancy by connector, waiting time, queue length, or real time station status. This would support models that separate price responses from congestion responses and quantify how pricing and queueing jointly shape participation.

A third direction is to pursue research designs that move beyond reduced form outcome responses toward clearer causal mechanisms. One approach is to study discrete tariff revisions, pricing rule changes, or staggered policy changes across operators and locations. Another approach is to evaluate combined interventions, such as power upgrades or targeted densification paired with pricing changes. This can help test whether infrastructure relaxes binding constraints and changes the main adjustment margin. Work on charging station congestion and network equilibrium also suggests that demand responses depend on network structure and congestion feedback, which supports combining behavioral analysis with explicit network and queueing models (Huang and Kockelman, 2020).

Finally, future work should treat pricing as one part of a broader charging system. Pricing interacts with infrastructure capacity, reliability, information provision, and user heterogeneity. Evidence on public charging equity and reliability indicates that access and performance can differ across communities, which can shape who uses public charging and how responsive they can be (Yu et al., 2025). Linking these elements with high frequency operational data would support more context aware policy design and help move from documenting outcome responses to explaining the mechanisms that generate them.

**Reference**

Bayram, I. S., & Galloway, S. (2021). Pricing-based distributed control of fast EV charging stations operating under cold weather. *IEEE Transactions on Transportation Electrification*, *8*(2), 2618-2628.

Borlaug, B., Yang, F., Pritchard, E., Wood, E., & Gonder, J. (2023). Public electric vehicle charging station utilization in the United States. *Transportation Research Part D: Transport and Environment*, *114*, 103564.

Daina, N., Sivakumar, A., & Polak, J. W. (2017). Electric vehicle charging choices: Modelling and implications for smart charging services. *Transportation Research Part C: Emerging Technologies*, *81*, 36-56.

Faruqui, A., & Sergici, S. (2010). Household response to dynamic pricing of electricity: a survey of 15 experiments. *Journal of regulatory Economics*, *38*(2), 193-225.

Funke, S. Á., Plötz, P., & Wietschel, M. (2019). Invest in fast-charging infrastructure or in longer battery ranges? A cost-efficiency comparison for Germany. *Applied energy*, *235*, 888-899.

Funke, S. Á., Sprei, F., Gnann, T., & Plötz, P. (2019). How much charging infrastructure do electric vehicles need? A review of the evidence and international comparison. *Transportation research part D: transport and environment*, *77*, 224-242.

Guo, Z., You, L., Zhu, R., Zhang, Y., & Yuen, C. (2025). A City-scale and Harmonized Dataset for Global Electric Vehicle Charging Demand Analysis. *Scientific Data*, *12*(1), 1254.

He, S. Y., Kuo, Y. H., & Sun, K. K. (2022). The spatial planning of public electric vehicle charging infrastructure in a high-density city using a contextualised location-allocation model. *Transportation Research Part A: Policy and Practice*, *160*, 21-44.

Hecht, C., Figgener, J., & Sauer, D. U. (2022). Analysis of electric vehicle charging station usage and profitability in Germany based on empirical data. *Iscience*, *25*(12).

Helmus, J. R., Spoelstra, J. C., Refa, N., Lees, M., & van den Hoed, R. (2018). Assessment of public charging infrastructure push and pull rollout strategies: The case of the Netherlands. *Energy Policy*, *121*, 35-47.

Huang, Y., & Kockelman, K. M. (2020). Electric vehicle charging station locations: Elastic demand, station congestion, and network equilibrium. *Transportation Research Part D: Transport and Environment*, *78*, 102179.

Jones, C. B., Vining, W., Lave, M., Haines, T., Neuman, C., Bennett, J., & Scoffield, D. R. (2022). Impact of Electric Vehicle customer response to Time-of-Use rates on distribution power grids. *Energy Reports*, *8*, 8225-8235.

Kristoffersson, I., Pyddoke, R., Kristofersson, F., & Algers, S. (2025). Access to charging infrastructure and the propensity to buy an electric car. *Transportation Research Part D: Transport and Environment*, *139*, 104588.

Kuang, H., Zhang, X., Qu, H., You, L., Zhu, R., & Li, J. (2024). Unraveling the effect of electricity price on electric vehicle charging behavior: A case study in Shenzhen, China. *Sustainable Cities and Society*, *115*, 105836.

Lee, Z. J., Li, T., & Low, S. H. (2019, June). ACN-data: Analysis and applications of an open EV charging dataset. In *Proceedings of the tenth ACM international conference on future energy systems* (pp. 139-149).

Lei, T., Guo, S., Qian, X., & Gong, L. (2022). Understanding charging dynamics of fully-electrified taxi services using large-scale trajectory data. *Transportation Research Part C: Emerging Technologies*, *143*, 103822.

Li, J., Chew, A., & Wang, H. (2024). Investigating state-of-the-art planning strategies for electric vehicle charging infrastructures in coupled transport and power networks: A comprehensive review. *Progress in Energy*.

Limmer, S. (2019). Dynamic pricing for electric vehicle charging—a literature review. *Energies*, *12*(18), 3574.

Liu, Y. S., Tayarani, M., & Gao, H. O. (2022). An activity-based travel and charging behavior model for simulating battery electric vehicle charging demand. *Energy*, *258*, 124938.

Luo, S., & Wang, Y. (2025). The impacts of rate surge on electric vehicle charging behaviors: Evidence from California. *Journal of Economic Behavior & Organization*, *236*, 107048.

Motoaki, Y., Yi, W., & Salisbury, S. (2018). Empirical analysis of electric vehicle fast charging under cold temperatures. *Energy Policy*, *122*, 162-168.

Potoglou, D., Song, R., & Santos, G. (2023). Public charging choices of electric vehicle users: A review and conceptual framework. *Transportation Research Part D: Transport and Environment*, *121*, 103824.

Powell, S., Cezar, G. V., Min, L., Azevedo, I. M., & Rajagopal, R. (2022). Charging infrastructure access and operation to reduce the grid impacts of deep electric vehicle adoption. *Nature Energy*, *7*(10), 932-945.

Schroeder, A., & Traber, T. (2012). The economics of fast charging infrastructure for electric vehicles. *Energy Policy*, *43*, 136-144.

Shahraki, N., Cai, H., Turkay, M., & Xu, M. (2015). Optimal locations of electric public charging stations using real world vehicle travel patterns. *Transportation Research Part D: Transport and Environment*, *41*, 165-176.

Shariatzadeh, M., Lopes, M. A., & Antunes, C. H. (2025). Electric vehicle users' charging behavior: A review of influential factors, methods and modeling approaches. *Applied Energy*, *396*, 126167.

Small, K. A. (1982). The scheduling of consumer activities: work trips. *The American Economic Review*, *72*(3), 467-479.

Springel, K. (2021). Network externality and subsidy structure in two-sided markets: Evidence from electric vehicle incentives. *American Economic Journal: Economic Policy*, *13*(4), 393-432.

Varshney, S., Panda, K.P., Shah, M., Srinivas, B.A., Deshmukh, A., Choudhary, K.K., Bajaj, M., Prokop, L. & Rubanenko, O. (2025). Novel control strategies for electric vehicle charging stations using stochastic modeling and queueing analysis. *Scientific Reports*, *15*(1), 21391.

Vickrey, W. S. (1969). Congestion theory and transport investment. *The American economic review*, *59*(2), 251-260.

Wolbertus, R., & Gerzon, B. (2018). Improving electric vehicle charging station efficiency through pricing. *Journal of advanced transportation*, *2018*(1), 4831951.

Wolbertus, R., Kroesen, M., Van Den Hoed, R., & Chorus, C. (2018). Fully charged: An empirical study into the factors that influence connection times at EV-charging stations. *Energy Policy*, *123*, 1-7.

Wolbertus, R., Kroesen, M., van den Hoed, R., & Chorus, C. G. (2018). Policy effects on charging behaviour of electric vehicle owners and on purchase intentions of prospective owners: Natural and stated choice experiments. *Transportation Research Part D: Transport and Environment*, *62*, 283-297.

Wolbertus, R., van den Hoed, R., Kroesen, M., & Chorus, C. (2021). Charging infrastructure roll-out strategies for large scale introduction of electric vehicles in urban areas: An agent-based simulation study. *Transportation Research Part A: Policy and Practice*, *148*, 262-285.

Yang, S. Y., Woo, J., & Lee, W. (2024). Assessing optimized time-of-use pricing for electric vehicle charging in deep vehicle-grid integration system. *Energy Economics*, *138*, 107852.

Yang, X., Peng, Z., Wang, P., & Zhuge, C. (2023). Seasonal variance in electric vehicle charging demand and its impacts on infrastructure deployment: A big data approach. *Energy*, *280*, 128230.

Yu, Q., Que, T., Cushing, L.J., Pierce, G., Shen, K., Kejriwal, M., Yao, Y. & Zhu, Y. (2025). Equity and reliability of public electric vehicle charging stations in the United States. *Nature Communications*, *16*(1), 5291.

Zhang, L., Shaffer, B., Brown, T., & Samuelsen, G. S. (2015). The optimization of DC fast charging deployment in California. *Applied energy*, *157*, 111-122.